\documentclass[a4paper,11pt]{article}
\pdfoutput=1 

\usepackage{jheppub}

\newcommand{\ii}{\mathrm{i}}
\newcommand{\eee}{\text{e}}
\newcommand{\tr}{\mathrm{tr}\,}
\newcommand{\pd}{\partial}
\newcommand{\one}{{\rm 1\kern -.9mm l}}
\newcommand{\p}{\partial}

\title{\boldmath S-duality, triangle groups and modular anomalies in $\mathcal{N}=2$ SQCD}

\author[a]{S. K. Ashok,}
\author[a]{E. Dell'Aquila,}
\author[b]{A. Lerda\,}
\author[a]{and M. Raman\,}

\affiliation[a]{Institute of Mathematical Sciences \\
   C. I. T. Campus, Taramani\\
   Chennai, India 600113}

\affiliation[b]{Universit\`a del Piemonte Orientale, Dipartimento di Scienze e Innovazione Tecnologica, \\
and I. N. F. N. - Gruppo Collegato di Alessandria - sezione di Torino\\
Viale T. Michel  11, I-15121 Alessandria, Italy}

\emailAdd{sashok@imsc.res.in}
\emailAdd{edellaquila@imsc.res.in}
\emailAdd{lerda@to.infn.it}
\emailAdd{madhur@imsc.res.in}

\abstract{
We study ${\mathcal N}=2$ superconformal theories with gauge group SU$(N)$ and $2N$ fundamental 
flavours in a locus of the Coulomb branch with a $\mathbb{Z}_N$ symmetry. In this special vacuum, we 
calculate the prepotential, the dual periods and the period matrix using equivariant localization. When the
flavors are massless, we find that the period matrix is completely specified by $\left[\frac{N}{2}\right]$ effective 
couplings. On each of these, we show that the S-duality group acts as a generalized triangle group 
and that its hauptmodul can be used to write a non-perturbatively exact relation between each 
effective coupling and the bare one. For $N=2, 3, 4$ and $6$, the generalized triangle group 
is an arithmetic Hecke group which contains a subgroup that is also a congruence subgroup of 
the modular group PSL(2,$\mathbb{Z}$). For these cases, we introduce mass deformations that respect the 
symmetries of the special vacuum and show that the constraints arising from S-duality make it possible to 
resum the instanton contributions to the period matrix in terms of meromorphic modular forms which solve 
modular anomaly equations.
}

\keywords{$\mathcal{N}=2$ SYM theories, S-duality, recursion relations, instantons}

\begin{document} 
\maketitle
\flushbottom

\section{Introduction}
\label{secn:intro}

One of the most interesting properties of supersymmetric gauge theories is the existence of 
non-perturbative S-dualities that relate their weak- and 
strong-coupling behaviour.\footnote{For a recent review we refer the 
reader to \cite{Teschner:2016yzf}.}
Recently, there has been much progress in understanding these 
dualities in conformally invariant $\mathcal{N}=2$ 
supersymmetric gauge theories in four dimensions, 
especially following the seminal work of Gaiotto \cite{Gaiotto:2009we}. 
In that work, the four-dimensional $\mathcal{N}=2$ theories were realized as 
compactifications of the six-dimensional $(2,0)$ theory on a punctured 
Riemann surface $\Sigma$. One of the important results of this approach was to identify 
the complex structure moduli space of $\Sigma$ with the space of gauge couplings modulo 
the action of the S-duality group. For linear quiver gauge theories in the weak coupling limit, the 
Riemann surface degenerates into a collection of three-punctured spheres connected by long thin tubes,
and the sewing parameters are identified with the bare coupling constants of the superconformal gauge theory.

This approach is fruitfully contrasted with the original solution of ${\mathcal N}=2$ gauge theories due to 
Seiberg and Witten \cite{Seiberg:1994rs,Seiberg:1994aj}, where the quantum effective action on the Coulomb 
branch is obtained from an algebraic curve describing a Riemann surface, and an associated holomorphic 
differential. 
For generic vacuum expectation values of the scalar fields in the adjoint gauge multiplet, the quantum
effective action describes a ${\mathcal N}=2$ supersymmetric theory with gauge group U$(1)^r$,
where $r$ is the rank of the original non-abelian gauge group. The matrix of effective couplings 
$\tau_{ij}$ between the various U(1)'s is identified with the period matrix of the 
Seiberg-Witten curve. 
For gauge groups with large $r$, it becomes difficult to use the Seiberg-Witten curve 
and the corresponding differential to do explicit calculations. In such cases, however, it is possible to make 
progress using equivariant localization methods
\cite{Nekrasov:2002qd,Nekrasov:2003rj,Bruzzo:2002xf,Fucito:2004ry}
which allow one to compute the prepotential, the dual periods and the period matrix of the effective action
order by order in an instanton expansion. Interestingly, the instanton counting parameters in this expansion
have a natural interpretation as the bare coupling constants of the superconformal gauge theory 
\cite{Gaiotto:2009we,Ashok:2015gfa}. 

Whichever approach one uses to study the low-energy theory, a natural 
question to ask is whether the non-perturbative S-duality group can be used to solve 
for the effective action. For
$\mathcal{N}=2^\star$ theories (i.e.~mass deformed $\mathcal{N}=4$ theories)
with unitary gauge groups it has been 
shown \cite{Minahan:1997if,Billo:2013jba,Billo:2013fi,Billo:2014bja} that the constraints 
coming from S-duality take the form of a modular anomaly equation
whose solution allows one to reconstruct the prepotential on the Coulomb branch
order by order in the mass of the adjoint hypermultiplet to all orders in the gauge coupling. 
To achieve this result one has to organize the low-energy effective 
prepotential as a semi-classical expansion in inverse powers of the vacuum 
expectation values of the scalar fields in the gauge vector multiplet 
and realize that the coefficients of this expansion
satisfy a recursion relation whose solution can be written in terms of quasi-modular forms 
of PSL(2,$\mathbb{Z}$) acting on the bare gauge coupling. 
These modular forms resum the instanton series and
therefore provide an exact result. It is of particular importance that ${\mathcal N}=2^\star$ theories are 
characterized 
by the absence of any renormalization of the coupling constant, even non-perturbatively; thus, the bare 
coupling is the only coupling that is present in the effective theory. 
This procedure has been applied also to $\mathcal{N}=2^\star$ theories with arbitrary gauge 
groups in \cite{Billo':2015jta,Billo':2015ria,Billo:2016zbf},  
where it has been observed that for non-simply laced algebras the effective prepotential is expressed 
in terms of quasi-modular forms of congruence subgroups of PSL(2,$\mathbb{Z}$).

In this work we study ${\mathcal N}=2$ gauge theories with gauge group SU$(N)$ and 
$2N$ fundamental flavours, generalizing the analysis of the $\text{SU}(3)$ 
gauge theory with six flavours recently presented in \cite{Ashok:2015cba}. 
When all flavours are massless, these SQCD theories are superconformal. However, 
unlike the case of $\mathcal{N}=2^\star$ theories, the bare gauge 
coupling in ${\mathcal N}=2$ SQCD is renormalized by quantum corrections which arise from a 
finite $1$-loop contribution as well as from an 
infinite series of non-perturbative contributions due to instantons. 
In general these corrections are different for the various U$(1)$ 
factors and thus one expects to find several effective couplings in the low-energy theory.

This paper is divided into two parts. In the first part, we work in the
limit where all flavours are massless, and calculate various observables of 
the effective theory such as the prepotential, the period integrals and the period matrix,  
using equivariant localization. In particular, we work in a special locus of the Coulomb branch  
which possesses a $\mathbb{Z}_N$ symmetry and which we call ``special vacuum'' \cite{Argyres:1999ty}. 
In this special vacuum, the period matrix has fewer independent components than it does at a 
generic point of the moduli space. More precisely, when all quantum corrections are taken into account  
there are $\left[\frac{N}{2}\right]$ distinct matrix structures which correspond to $\left[\frac{N}{2}\right]$
renormalized coupling constants in the effective theory.\footnote{Here $\left[\,\cdot\,\right]$ denotes the 
floor function.} Of course, at leading order 
such renormalized couplings are all equal to the bare coupling, but when $1$-loop and 
instanton corrections are taken into account, they begin to differ from one another. 
Given that the S-duality group naturally acts
on the bare coupling, an obvious question to ask is how S-duality is realized on the various
parameters of the quantum theory. 
The answer we provide in this paper is that on each individual effective coupling 
S-duality acts as a generalized triangle group (see for example \cite{Doran:2013npa}). 
Moreover, using this insight, we propose a non-perturbatively \emph{exact} relation between the bare 
coupling and the renormalized ones that takes a universal form in terms of the 
$j$-invariants of the triangle groups. We perform several successful checks of this
proposal by comparing the instanton contributions predicted by the exact relation with the explicit results 
obtained from multi-instanton localization.
As a further evidence in favour of our proposal, we show that the action of S-duality 
on the renormalized couplings is fully consistent with the action on the bare coupling as obtained 
from Gaiotto's analysis \cite{Gaiotto:2009we}. We believe that our results, and in particular 
the exact relation we propose, can play an important role in the study of these SQCD theories at strong coupling 
\cite{Argyres:2007cn}. This is because the $j$-invariants have a well-understood behaviour near those 
cusp points where the coupling constants become large and the usual weak-coupling expansion cannot be used.

In the second part of the paper we consider the case where the fundamental
flavour hypermultiplets are massive. For generic masses the $\mathbb{Z}_N$ symmetry of the special vacuum is broken; 
to avoid this, we restrict our analysis to $\mathbb{Z}_N$-symmetric mass configurations 
so that the modular structure uncovered in the massless limit gets deformed in a natural and smooth
manner. In particular, with these $\mathbb{Z}_N$-symmetric mass configurations 
we find that the $\left[\frac{N}{2}\right]$ matrix structures of the massless theories are preserved, while the 
$\left[\frac{N}{2}\right]$ effective couplings simply receive further contributions proportional 
to the hypermultiplet masses. Building on earlier literature \cite{Huang:2011qx,Billo':2011pr}, this analysis 
was already carried out for the SU(2) theory in \cite{Billo:2013jba,Billo:2013fi}, 
where it was shown that the prepotential can be written in terms of quasi-modular forms of the 
modular group PSL$(2,\mathbb{Z})$. Moreover, after expanding the prepotential in powers of the 
flavour masses, it was realized that the coefficients of this expansion satisfy a modular anomaly equation
that takes the form of a recursion relation, similar to that of the $\mathcal{N}=2^\star$ case. 
These results have been recently extended to the SU$(3)$ theory with
six massive flavours in \cite{Ashok:2015cba}, where it has been shown that
the prepotential, the dual periods and the period matrix are constrained by S-duality to obey again
a recursion relation that can be written as a modular anomaly equation.
In this case, the solutions of this equation are quasi-modular forms of $\Gamma_1(3)$, which is a 
subgroup of the S-duality group that is also a congruence subgroup of 
$\text{PSL}(2,\mathbb{Z})$.\footnote{The relevance of $\Gamma_1(3)$ and of its modular forms for the 
effective SU(3) theory with six flavours
was already observed long ago in \cite{Minahan:1995er,Minahan:1996ws,Minahan:1997fi}.} Here we 
further extend these results to the general SU$(N)$ theory with $2N$ massive flavours and show that
the constraints arising from S-duality can always be written as a recursion relation for any $N$. 
However, beyond this step, the analysis crucially depends on the arithmetic properties 
of the S-duality group. It turns out that for $N=2,3,4$ and $6$, the S-duality group acting on each 
quantum coupling always has a subgroup which is a congruence 
subgroup of $\text{PSL}(2,\mathbb{Z})$.
For these theories, which we call arithmetic, the discussion proceeds along the same 
lines described in \cite{Ashok:2015cba} for the SU$(3)$ theory, with one important modification: 
in the higher rank cases, the S-duality constraints are
written as \emph{coupled} modular anomaly equations. These 
coupled equations are nevertheless integrable and their solutions turn out to be polynomials in 
meromorphic quasi-modular forms of congruence subgroups of $\text{PSL}(2,\mathbb{Z})$.
For all non-arithmetic theories, instead, S-duality acts as generalized triangle groups and one would need 
to use their automorphic forms to solve for various observables. Here, we restrict our analysis only to the 
massive arithmetic cases, leaving the study of the non-arithmetic cases for the future. 

The paper ends with a discussion of the results and some future directions for work, and with two technical 
appendices.

\part{\boldmath Massless $\mathcal N=2$ SQCD and duality groups}
In this part, we discuss $\mathcal{N}=2$ SQCD theories with massless fundamental hypermultiplets.

\section{Massless superconformal QCD theories and special vacuum}
\label{secn:massless}
We begin by reviewing the main features of $\mathcal{N}=2$ SQCD theories with unitary gauge groups 
$\text{U}(N)$. These theories are superconformal invariant if the number of flavours is $2N$.

As usual, we can combine the bare Yang-Mills coupling $g$ and the $\theta$-angle into the complex variable
\begin{equation}
\tau_0 = \frac{\theta}{2\pi} + \ii\, \frac{4\pi}{g^2}~,
\label{tau0}
\end{equation}
so that the instanton counting parameter $q_0$ is defined as
\begin{equation}
q_0 = \eee^{2 \pi \ii \tau_0} ~.
\end{equation}
The low-energy effective dynamics of these $\mathcal{N}=2$ theories is completely determined by the 
prepotential, which we now describe.

\subsection{The prepotential}
The prepotential $F$ admits a decomposition into classical (tree-level), perturbative ($1$-loop), and non-
perturbative (instanton) contributions:
\begin{equation}
F=F_{\text{class}}+F_{\text{1-loop}}+F_{\text{inst}}~.
\label{prep}
\end{equation}

\subsubsection{Classical contribution}
For the U($N$) gauge theory the classical prepotential is given by
\begin{equation}
F_{\text{class}} = \ii \pi \tau_0 \,\tr\langle A\rangle^2 \,=\,\ii \pi \tau_0 \sum_{u=1}^{N} A_u^2 \, ,
\label{Fclass}
\end{equation}
where the vacuum expectation value of the adjoint scalar $A$ is
\begin{equation}
\langle A \rangle = \text{diag} \left( A_1, \cdots, A_{N} \right)~.
\end{equation}
For unitary gauge groups the $A_u$'s are unrestricted, while for special unitary groups we have to 
impose the tracelessness condition
\begin{equation}
\sum_{u=1}^{N} A_u =0~.
\label{traceless}
\end{equation}
Throughout this paper we satisfy this constraint by taking
\begin{align}
A_u = 
\begin{cases}
\qquad \ \ \quad a_u &\mbox{for} \quad u =1, \cdots, N-1~, \\
-\left( a_1 + \cdots + a_{N-1} \right) &\mbox{for} \quad u = N~.
\end{cases}
\label{suN}
\end{align}
When referring to the SU($N$) theory we will use the indices $i,j,\cdots \in \lbrace 1,\cdots,N-1\rbrace$
to label the Cartan directions. 

\subsubsection{Perturbative contribution}
The perturbative ($1$-loop) contribution to the prepotential is independent of the bare 
coupling $\tau_0$ and is given by
\begin{equation}
F_{\text{$1$-loop}} = \sum_{u \neq v=1}^N \gamma(A_u - A_v) 
- 2N\sum_{u=1}^{N} \gamma(A_u )~,
\label{F1loop}
\end{equation}
where (see for example \cite{D'Hoker:1999ft})
\begin{equation}
\gamma(x) = -\frac{x^2}{4} \log \left( \frac{x^2}{\Lambda^2} \right)~.
\end{equation}
Here $\Lambda$ is an arbitrary mass scale, which actually drops out from  $F_{\text{$1$-loop}}$ due to 
conformal invariance.

\subsubsection{Instanton contribution}
The non-perturbative contributions to the prepotential can be explicitly calculated using the methods of 
equivariant localization \cite{Nekrasov:2002qd,Nekrasov:2003rj,Bruzzo:2002xf,Fucito:2004ry} 
(see also \cite{Billo:2012st} for technical details) and are of the form
\begin{equation}
F_{\text{inst.}} = \sum_{k=1}^{\infty} F_k(u_r)\,q_0^k
\label{Finst}
\end{equation}
where 
\begin{equation}
u_r = \sum_{u=1}^{N} A_u^{\,r}
\label{ur}
\end{equation}
for $r=1,\cdots,N$ are the Casimir invariants of the gauge group.
The function $F_k$ represents the $k$-instanton contribution to the prepotential and, on dimensional grounds, 
must have mass dimension $2$. 

\subsection{The special vacuum}
In the following we will study the massless SQCD theories in the
so-called \emph{special vacuum} \cite{Argyres:1999ty} which is defined as the locus of points 
on the moduli space where 
\begin{equation}
u_r= 0 \quad \text{for} \quad r=1 , \cdots, N-1~.
\label{specialvacuum}
\end{equation}
For SU($N$) theories the condition $u_1=0$ is nothing but (\ref{suN}), while the other conditions 
select vacuum configurations with special properties.\footnote{In the
SU(2) theory there is clearly only one condition, namely $u_1=0$ and the notion of special vacuum does not 
apply in this case. Despite this fact, most of the subsequent formulas formally hold also for SU(2).
\label{footnoteSU2}}

The special vacuum restriction (\ref{specialvacuum}) can be implemented by choosing the
vacuum expectation values of the adjoint SU($N$) scalar as
\begin{equation}
a_i = \omega^{i-1} a
\label{aspecvac}
\end{equation}
for $i=1,\cdots,N-1$, where
\begin{equation}
\omega = \eee^{\frac{2\pi \ii}{N}}~.
\label{omega}
\end{equation} 
We thus see that the special vacuum can be parametrized by a single scale $a$ and that it possesses 
a $\mathbb{Z}_N$ symmetry.

\subsection{Observables in the special vacuum}
\label{subsecn:observ}
We now discuss the properties of some observables in the special vacuum, starting with the prepotential.

\subsubsection{The prepotential}
In the special vacuum several simplifications occur when one evaluates the prepotential. For instance, the 
classical  prepotential (\ref{Fclass}) vanishes and the $\mathbb{Z}_N$-invariance of the special vacuum implies 
that for large $a$ the prepotential has a semi-classical expansion of the form
\begin{equation}
\label{Fsv}
F = \sum_{n=1}^{\infty} \frac{f_n \big(q_0 \big)}{a^{n N}}~.
\end{equation}
The coefficients $f_n$'s must have mass dimension equal to $(n N+2)$; however, since the flavours are 
massless, the only available scale is $a$ and it is not possible to give $f_n$ the required mass dimensions. 
Thus the prepotential identically vanishes in the special vacuum.\footnote{The case $N=2$ is clearly an 
exception. Indeed, the prepotential of the massless SU$(2)$ theory is 
proportional to $a^2$, which has the right mass dimension and is $\mathbb{Z}_2$-symmetric (see for instance 
\cite{Billo:2013fi, Billo:2013jba}).}

\subsubsection{Dual periods}
In the SU$(N)$ theory the dual periods $a^{\text D}_{i}$ are defined by
\begin{equation}
a^{\text{D}}_{i} = \frac{1}{2\pi \ii}\frac{\partial F}{\partial a_i} ~.
\label{adual}
\end{equation}
As in the special vacuum all $a_i$'s are proportional to each other, 
this is also true of the dual periods. For example one can verify that
\begin{equation}
a^{\text{D}}_{i}=-\big(\omega+\omega^2+\cdots+\omega^i\big)a^{\text{D}}_{N-1}
\label{adualprop}
\end{equation}
for any $i=1,\cdots,N-1$. Therefore, in the special vacuum without any loss of generality
we can choose the following conjugate pair of variables: $(a^{\text{D}}_{N-1}, a_1)$. 
To simplify notation, we will omit the subscripts and denote these just by $(a^{\text{D}}, a)$. 

The classical contribution to $a^{\text{D}}$ is given by
\begin{align}
a^{\text{D}}_{\text{class}} &=  \tau_0 \left(a_1 + a_2 +\cdots +2 a_{N-1} \right) \cr 
&= c_N\, \tau_0\, a~,
\label{adclass}
\end{align}
where the second line follows upon using the special vacuum values (\ref{aspecvac}) which lead to
\begin{align}
c_N= \frac{(1-\omega)}{\omega^2} ~.
\label{cNgeneral}
\end{align}
The classical dual period receives both $1$-loop and instanton corrections, even in the massless theory. 
Physically, this corresponds to a non-perturbative redefinition of the bare coupling constant $\tau_0$ into a 
new renormalized coupling constant that we denote $\tau$. This renormalized coupling constant is defined in 
such a way that the quantum corrected dual period takes the simple form
\begin{equation}
\label{admassless}
a^{\text{D}} = c_N\, \tau\, a~,
\end{equation}
namely the same classical expression (\ref{adclass}) with $\tau_0$ replaced by $\tau$. The latter admits the 
following non-perturbative expansion
\begin{equation}
2\pi \ii\,\tau= 2\pi \ii\,\tau_0 + \ii \pi + \log  b_0   + \sum_{k=1}^{\infty} b_k\, q_0^k ~.
\label{tauexp}
\end{equation}
In this expression, the logarithmic term represents a finite contribution at $1$-loop, while the 
term proportional to $q_0^k$ is the $k$-instanton contribution.

\subsubsection{The period matrix}
In the SU($N$) theory the period matrix $\Omega$ is the $(N-1)\times (N-1)$ matrix defined as
\begin{equation}
\Omega_{ij} = \frac{1}{2\pi \ii} \frac{\pd^2 F}{\pd a_i \pd a_j}~.
\label{Omegaij}
\end{equation}
The classical part of the period matrix is simply given by
\begin{equation}
\label{Omegaclass}
\Omega_{\text{class}} = \tau_0\,\mathcal{C}
\end{equation}
where
\begin{equation}
\mathcal{C}=\begin{pmatrix}
 ~2 & ~1~ & \cdots & 1~ \\
 ~1 & ~2~ & \cdots & 1~ \\
 \vdots & \vdots & \ddots & \vdots \\
 ~1 & ~1~ & \hdots & 2~
\end{pmatrix}
\label{Cartan}
\end{equation}
is the Cartan matrix corresponding to our parametrization of SU$(N)$. 

For $N>3$ this simple structure is lost \cite{Minahan:1996ws,AhaYan96} when perturbative and instanton 
contributions are taken into account, even in the special vacuum. 
For example, at $1$-loop from (\ref{F1loop}) one finds
\begin{equation}
\label{Omega1loop}
\Omega_{\text{1-loop}} = \frac{\ii}{\pi}\,\Big(\log(2N)\, \mathcal{C}+\mathcal{G}\Big)
\end{equation}
where the matrix elements of $\mathcal{G}$ are given by \cite{Minahan:1996ws}
\begin{equation}
\begin{aligned}
\mathcal{G}_{ii}& = 2\log\sin\left( \frac{i \pi}{N} \right) ~,\\
\mathcal{G}_{ij}&=\log\sin\left( \frac{i \pi}{N} \right)+ \log\sin\left( \frac{j \pi}{N} \right)- 
\log\sin\left( \frac{|i-j| \pi}{N} \right)\quad\mbox{for}~i\not=j~.
\end{aligned}
\label{G}
\end{equation}
For $N=3$ it is easy to see that $\mathcal{G}$ is proportional to the Cartan matrix $\mathcal{C}$, but this
relation does not hold for $N>3$. Indeed, a closer inspection of (\ref{G}) reveals that it is possible to identify
$\left[\frac{N}{2}\right]$ different matrix structures.
A similar result is found even after the instanton contributions are taken into account. Thus, in general the 
complete period matrix $\Omega$ can be written as
\begin{equation}
\Omega = \underbrace{\tau_{1} \ \mathcal{M}_1 + \tau_{2} \ \mathcal{M}_2 
+ \cdots\phantom{\big|}}_{\big[\frac{N}{2}\big]\,\text{terms}}
\label{Omegafull}
\end{equation}
where the ${\cal M}_k$'s are independent matrix structures and the $\tau_k$ are distinct complex couplings 
that characterize the effective theory.
Of course, one could in principle use any basis of matrices $\mathcal{M}_k$ to write $\Omega$, but a 
particularly insightful choice is the one that ``diagonalizes'' the action of the S-duality group. 
In such a basis, under
S-duality each $\mathcal{M}_k$ stays invariant and each $\tau_k$ transforms \emph{individually} as
\begin{equation}\label{tautildeexp}
\tau_k \rightarrow - \frac{1}{\lambda_k \,\tau_k}
\end{equation}
for some positive $\lambda_k$. We will explicitly show in a series of examples that the 
spectrum of $\lambda_k$ is given by
\begin{equation}
\lambda_k = 4 \sin^2 \left( \frac{k\, \pi}{N} \right) ~.
\label{lambdak}
\end{equation}
Note that for $N\in \{2,3,4,6\}$ all the $\lambda_k$'s take integer values. We call these cases 
\emph{arithmetic}. If instead $N \not \in \{2,3,4,6\}$, then the $\lambda_k$'s are not necessarily integer. We 
refer to the latter as the \emph{non-arithmetic} cases.
Moreover, we will find that for any $N$ the coupling $\tau_1$ in (\ref{Omegafull}) 
coincides with the coupling $\tau$ that appears in the expression (\ref{admassless}) for 
the dual period $a^{\text{D}}$.

In order to show these facts, we now turn to a detailed discussion of the S-duality group.

\section{The S-duality group}
\label{Sduality}

The S-duality group of $\mathcal{N}=2$ SQCD has been derived in \cite{Argyres:1998bn}. Here, we focus on 
the massless case in the special vacuum, for which the Seiberg-Witten curve takes the following hyperelliptic 
form
\begin{equation} 
y^2 = (x^N-u_N)^2 - h\, x^{2N} ~.
\label{SW}
\end{equation}
Here $u_N$ is the only non-zero Coulomb modulus labeling the special vacuum 
and $h$ is a function of the gauge coupling given by (see for example \cite{Billo:2012st, Ashok:2015cba})%
\begin{equation}
h=\frac{4q_0}{(1+q_0)^2}~.
\label{hq0}
\end{equation}
The Seiberg-Witten curve degenerates when its discriminant vanishes and from (\ref{SW}) it is easy to see
that this happens at $h=0,1, \infty$. The monodromies around these points generate the S-duality group 
\cite{Argyres:1998bn}. We will take this to be our working definition of the S-duality group in what follows. In 
Section \ref{Saction} we will rederive this result by a completely different method.

We begin by choosing a canonical homology basis of cycles for the U$(N)$ theory described by (\ref{SW}), 
which we denote by hatted variables. Specifically, we introduce $\hat{\alpha}$ and $\hat\beta$ cycles 
with the following intersections
\begin{equation}
\begin{aligned}
\hat\alpha_u \cap \hat\alpha_v &= \hat\beta_u \cap \hat\beta_v = 0\phantom{\Big|}~,\\
\Big(\hat\alpha_u \cap \hat\beta_v \Big)&= 
\begin{pmatrix}
1 & -1 & 0 & \cdots & 0 \\
0 & 1 & -1 & \cdots & 0 \\
\vdots & \vdots & \vdots & \ddots & \vdots \\
-1 & 0 & 0 & \cdots & 1
\end{pmatrix}~,
\end{aligned}
\label{intersect}
\end{equation}
for $u,v=1,\cdots,N$.
These cycles are not linearly independent since
\begin{equation}
\sum_{u=1}^N\hat\alpha_u =0\qquad\mbox{and}\qquad\sum_{v=1}^N\hat\beta_v =0~.
\label{relationab}
\end{equation}
In the special vacuum there is a natural $\mathbb{Z}_N$ symmetry that rotates this basis clockwise and is 
generated by
\begin{align}
\Phi \quad : \quad \begin{cases}
\hat\alpha_u &\longrightarrow \ \hat\alpha_{u-1} ~, \\
\hat\beta_v &\longrightarrow \ \hat\beta_{v-1} ~.
\end{cases}
\label{zninv}
\end{align}
Physical observables are insensitive to this $\mathbb{Z}_N$ rotation.

The action of $S$- and $T$-transformations on this basis of cycles has been determined in 
\cite{Argyres:1998bn} from the monodromy around the points $h=\infty$ and $h=0$, respectively, 
and is given by
\begin{equation}
S \quad : \quad \begin{cases}
\hat\alpha_u \rightarrow \hat\beta_{u} ~, \\
\hat\beta_v \rightarrow \hat\alpha_{v-1} ~,
\end{cases}\qquad\mbox{and}
\qquad
T \quad : \quad \begin{cases}
\hat\alpha_u \rightarrow \hat\alpha_{u}~, \\
\hat\beta_v \rightarrow \hat\beta_{v}+\hat\alpha_{v}-\hat\alpha_{v-1} ~.
\end{cases}
\label{STtransf}
\end{equation}
In the SU($N$) theory we can choose the independent cycles as follows:
\begin{equation}
\alpha_i = \hat{\alpha}_i \qquad\mbox{and}\qquad
\beta_j = \sum_{i=1}^{j} \hat{\beta}_i 
\label{suncycles}
\end{equation}
for $i,j=1,\cdots,N-1$. Using (\ref{intersect}) one can easily check that this basis is symplectic, 
in the sense that $\alpha_i \cap \alpha_j = \beta_i \cap \beta_j =0$ and 
$\alpha_i \cap \beta_j = \delta_{ij}$.

The restriction of the $S$ and $T$ transformations to the SU($N$) basis (\ref{suncycles}) follows directly
from (\ref{STtransf}). If we represent them as $(2N-2)\times (2N-2)$
matrices acting on the $(2N-2)$ vector {\small{$\begin{pmatrix}\beta\\ \alpha \end{pmatrix}$}}, we find
\begin{equation}
S =\begin{pmatrix}
0 & ~\mathcal{B}~\\
-(\mathcal{B}^{\,\text{t}})^{-1} & ~0~
\end{pmatrix}\qquad
\mbox{and}
\qquad
T =\begin{pmatrix}
~\one~& ~\mathcal{C}~\\
0 & \one
\end{pmatrix}
\label{STmatrices}
\end{equation}
where
\begin{equation}
\mathcal{B} = \begin{pmatrix}
-1 & -1 & -1 & \cdots  & -1  \\
0 &  -1 & -1 & \cdots & -1 \\
0 & 0 & -1 & \cdots &  -1 \\
\vdots & \vdots & \vdots & \ddots & \vdots \\
0 & 0 & 0 & \cdots  & -1
\end{pmatrix}~,
\label{Bmatrix}
\end{equation}
and $\mathcal{C}$ is the Cartan matrix (\ref{Cartan}). It is interesting to observe that 
\begin{equation}
S^2 = V~,
\end{equation}
where $V$ is an Sp$(2N-2, \mathbb{Z})$ matrix that implements the $\mathbb{Z}_N$ transformation
(\ref{zninv}) on the SU$(N)$ basis of cycles $(\alpha_i, \beta_j)$, given by
\begin{equation}
{V} = \begin{pmatrix}
(\mathcal{V}^{\,\text{t}})^{-1} & 0 \\
0 &  \mathcal{V}
\end{pmatrix}
\end{equation}
where
\begin{equation}
\mathcal{V}= \begin{pmatrix}
-1 & -1 & -1 & \cdots& -1\\
1 & 0   & 0 & \cdots & 0\\
0 & 1  & 0 &\cdots& 0\\
\vdots & \vdots & \ddots &\vdots & \vdots \\
0 & 0 & \cdots & 1 & 0  
\end{pmatrix}~.
\end{equation}
Notice that $\mathcal{V}^N=1$. Such a transformation leaves the period matrix 
invariant and is a symmetry of the theory. This means that $S$ effectively squares to the identity. 

\subsection{$S$-action on period integrals and gauge coupling}

{From} the transformations (\ref{STmatrices}) on the homology cycles we can straightforwardly deduce how
$S$ and $T$ act on the periods $a_i$ and their duals $a^{\text{D}}_j$, which are the integrals of the
Seiberg-Witten differential associated to the curve (\ref{SW}) over the cycles $\alpha_i$ and $\beta_j$ 
respectively.
Focusing in particular on the $S$-transformation, we find
\begin{equation}
\begin{aligned}
S(a^{\text{D}}_j) &= \big(\mathcal{B}\cdot a\big)_j~,\\
S(a_i) &= -\big((\mathcal{B}^{\,\text{t}})^{-1}\cdot a^{\text{D}}\big)_i
\end{aligned}
\end{equation}
where $\mathcal{B}$ is the matrix in (\ref{Bmatrix}). Thus, for our conjugate pair of 
variables $(a^{\text{D}}_{N-1},a_1)\equiv (a^{\text{D}}, a)$ we have
\begin{equation}
\begin{aligned}
S(a^{\text{D}}) &= -a_{N-1} = -\frac{1}{\omega^2}\, a~,\\
S(a) &= a_1^{\text{D}}=-\omega\,a^{\text{D}}~,
\end{aligned}
\label{saadual}
\end{equation}
where in each line the second equality follows upon using the special vacuum relations (\ref{aspecvac}) and
(\ref{adualprop}). As expected, the period and dual period integrals are exchanged under S-duality.

This result is quite useful since it allows us to deduce how S-duality acts on the gauge coupling $\tau$. 
Indeed, if we take into account the link (\ref{admassless}) between $a^{\text{D}}$ and $a$, and apply to it
the S-duality transformations, we find
\begin{equation}
S(a^{\text{D}}) =-c_N\,S(\tau)\,\omega\,a^{\text{D}}= -c_N^2 \,S(\tau)\,\tau\,\omega\,a~.
\label{Sad1}
\end{equation}
Consistency with (\ref{saadual}) implies that 
\begin{equation}
\label{stau}
\tau \rightarrow -\frac{1}{\lambda\,\tau}
\end{equation}
with
\begin{equation}
\lambda=-c_N^2\,\omega^3=-\frac{(1-\omega)^2}{\omega}=4 \sin^2 \frac{\pi}{N} ~.
\label{lambda10}
\end{equation}
Here we have used (\ref{cNgeneral}) and (\ref{omega}). This is precisely the case $k=1$ of the general
formula (\ref{lambdak}), and thus we conclude that the coupling $\tau$ that appears in the relation
between $a$ and $a^{\text{D}}$ is actually $\tau_1$, according to our definition in (\ref{tautildeexp}).
As we have already noticed, in the arithmetic cases $N \in \lbrace 2,3,4,6 \rbrace$, the constant $\lambda$ 
in (\ref{lambda10}) takes integer values.

\section{The arithmetic theories}
\label{sec:arithmeticity}

In this section, we collect the results obtained from localization calculations for the lower rank SQCD models and 
accumulate evidence for our conjecture regarding the form of the period matrix and the S-duality 
transformations of the quantum couplings that we anticipated at the end of Section~\ref{subsecn:observ}. 
While the SU$(2)$ and SU$(3)$ theories have already been studied in the literature, for completeness we
start by briefly recalling the main results for these cases.

\subsection{$N=2$}

In this case, the period matrix is just a complex constant given by
\begin{equation}
\Omega = 2\tau_1
\end{equation}
where $\tau_1$ is the only effective coupling of this theory.
Using multi-instanton calculations (see for example \cite{Billo:2012st,Ashok:2015gfa}), one can show that
\begin{equation}
\label{tau1vsq0}
2\pi\ii \, \tau_1 = \log q_0 + \ii \pi -\log16 + \frac{1}{2}\,q_0 + \frac{13}{64}\, q_0^2 
+\frac{23}{192} \,q_0^3 \cdots \, ,
\end{equation}
which can be inverted order by order to give
\begin{equation}
q_0 = - 16 \, q_1 \, (1+8\,q_1+44\,q_1^2+\cdots) = - 16\left(\frac{\eta(4\tau_1)}{\eta(\tau_1)}\right)^{8}~, 
\label{q0SU2}
\end{equation}
where $q_1 = \eee^{2\pi i \tau_1}$ and $\eta$ is the Dedekind $\eta$-function. The non-trivial relation between
$q_0$ and the effective coupling for the SU(2) theory was first noticed in \cite{Dorey:1996bn,Grimm:2007tm}.

The analysis of the previous section shows that under the $S$-transformation, the period matrix transforms 
under a symplectic Sp$(2,\mathbb{Z})$ 
transformation\,\footnote{See section \ref{SU2special} for a more detailed discussion of S-duality for 
the SU$(2)$ theory.}:
\begin{equation}
S\quad:\quad \Omega\rightarrow -\,\frac{1}{\Omega}~.
\end{equation}
{From} this it follows that the effective coupling $\tau_1$ transforms as
\begin{equation}
S\quad:\quad \tau_1\rightarrow -\frac{1}{4\tau_1} \, ,
\label{Stau1su2}
\end{equation}
in agreement with (\ref{tautildeexp}) since $\lambda_1=4$ for $N=2$.
Using this in (\ref{q0SU2}), we get
\begin{equation}
S\quad:\quad q_0 \rightarrow \frac{1}{q_0} ~.
\end{equation}
Furthermore, by computing the dual period we find
\begin{equation}
a^{\text{D}}= 2\tau_1 a
\label{adualSU2}
\end{equation}
in agreement with the general formula (\ref{admassless}) for $\omega=-1$.

\subsection{$N=3$}
In this case the period matrix turns out to be proportional to the SU$(3)$ Cartan matrix 
\begin{equation}
\Omega = \tau_1\begin{pmatrix}
2 &~1 \\
1&~2\\
\end{pmatrix}
\end{equation}
where
$\tau_1$ has the following instanton expansion (see for example \cite{Billo:2012st, Ashok:2015cba})
\begin{equation}
2\pi\ii \, \tau_1 = \log q_0 + \ii \pi -\log27 + \frac{4}{9}\,q_0 
+ \frac{14}{81}\, q_0^2 +\frac{1948}{19683}\,q_0^3 \cdots~.
\label{q0tausu3}
\end{equation}
As for the SU$(2)$ case, the SU$(3)$ theory in the special vacuum has a single $\tau_1$-parameter even after 
the quantum corrections are taken into account. On inverting the above expansion, we get
\begin{equation}
q_0 = - 27 \, q_1 \, (1+12\,q_1+90\,q_1^2+\cdots) = -27\left(\frac{\eta(3\tau_1)}{\eta(\tau_1)}\right)^{12}
\label{eta3}
\end{equation}
where, as before, $q_1=\eee^{2\pi\ii\tau_1}$.
Again, we have provided a non-perturbatively exact expression in terms of $\eta$-quotients.
Using the Sp$(4,\mathbb{Z})$ matrices derived in Section~\ref{Sduality}, one can check that S-duality 
leaves the SU$(3)$ Cartan matrix invariant and acts on $\tau_1$ as \cite{Ashok:2015cba}:
\begin{equation}
S\quad:\quad \tau_1 \rightarrow -\frac{1}{3\tau_1}~.
\end{equation}
in agreement with (\ref{lambdak}) since $\lambda_1=3$ for $N=3$; using this in (\ref{eta3}), we easily see
again that
\begin{equation}
S\quad:\quad q_0 \rightarrow \frac{1}{q_0} ~.
\end{equation}
Finally, on computing the dual period in this case we find
\begin{equation}
a^{\text{D}}= \ii \sqrt{3}\tau_1 a
\label{adualSU3}
\end{equation}
which confirms the general formula (\ref{admassless}) since for SU(3) $\omega=\eee^{\frac{2\pi\ii}{3}}$.

\subsection{$N=4$}

We now turn to the SU$(4)$ theory. As always, the classical period matrix is proportional to the Cartan matrix 
of the gauge Lie algebra but this time another independent matrix structure appears when one takes into 
account the 1-loop and the instanton corrections.
We have explicitly checked up to three instantons that it is possible to write the quantum period matrix as 
\begin{equation}
\Omega = \tau_1\, \mathcal{M}_1 + \tau_2\, \mathcal{M}_2 ~,
\label{omega4}
\end{equation}
where $\mathcal{M}_1$ and $\mathcal{M}_2$ are two $3 \times 3$ matrices given by
\begin{equation}
\label{M2M4matrices}
\mathcal{M}_1 = \begin{pmatrix}
\,1 \,& \,1\, &\, 0 \,\\
\,1 &\, 2\, & \,1\, \\
\,0\, & \,1\, & \,1\,
\end{pmatrix} \quad \text{and} \quad \mathcal{M}_2 = \begin{pmatrix}
\,1\, & \,0\, & \,1\, \\
\,0\, &\, 0\, & \,0\, \\
\,1\, & \,0\, & \,1\,
\end{pmatrix} ~,
\end{equation}
and the two couplings $\tau_1$ and $\tau_2$ have the following instanton expansions
\begin{subequations}
\begin{align}
2\pi\ii\,\tau_1 &= \log q_0 + \ii \pi -\log 64 + \frac{3 }{8}\,q_0
+\frac{141 }{1024}\,q_0^2+\frac{311}{4096}\, q_0^3 + \cdots ~,\label{tau1SU4}
\\
2\pi\ii\,\tau_2 &= \log q_0 + \ii \pi -\log 16 + \frac{1}{2}\,q_0+\frac{13 }{64}\,q_0^2
+\frac{23}{192}\,q_0^3 + \cdots ~.\label{tau2SU4}
\end{align}
\label{uvirmapsu4}
\end{subequations}
On inverting these expansions we find
\begin{subequations}
\begin{align}
q_0 &= - 64 \, q_1 \, (1+24\, q_1+300 \,q_1^2 + \cdots)  = 
- 64\left(\frac{\eta(2\tau_1)}{\eta(\tau_1)}\right)^{24}~, \label{q0qkSU41}\\
q_0 &= - 16 \, q_2 \, (1+8 \,q_2+44 \,q_2^2 + \cdots)= 
- 16\left(\frac{\eta(4\tau_2)}{\eta(\tau_2)}\right)^{8}~,
\label{q0qkSU42}
\end{align}
\label{q0qkSU4}
\end{subequations}
where we have introduced the notation
\begin{equation}
q_k = \eee^{2\pi \ii \tau_k}
\label{qk}
\end{equation}
for $k=1,2$. Once again, as for $N=2$ and 3, the bare coupling can be expressed as a quotient of $\eta$-
functions of the renormalized couplings. Notice that the $q_0$-expansion of $\tau_2$ is the same as that of 
the effective coupling 
of the SU(2) theory (see (\ref{tau1vsq0})). Although this coincidence may appear surprising at first glance, 
it is actually a consequence of the fact that this pair of couplings transform in the same way under S-duality. We 
will explicitly show this below, but this result can be anticipated by noticing 
that the general formula (\ref{lambdak}) implies that
$\lambda_2$ for $N=4$ and $\lambda_1$ for $N=2$ are both equal to 4.\footnote{Indeed, for all even 
$N$, $\lambda_{\frac{N}{2}} = 4$.}

Let us now consider the action of S-duality on the period matrix (\ref{omega4}). Using the 
Sp$(6,\mathbb{Z})$ transformations (\ref{STmatrices}), we find that 
$\mathcal{M}_1$ and $\mathcal{M}_2$ are
left invariant while
\begin{equation}
\label{Sontau_k}
S\quad:\quad \tau_k \rightarrow -\frac{1}{\lambda_k\,\tau_k} 
\end{equation}
with $\lambda_1 =2$ and $\lambda_2=4$, exactly as predicted by (\ref{tautildeexp}) and (\ref{lambdak}). 
Using these transformations in (\ref{q0qkSU4}), we can check also in this case that
\begin{equation}
S\quad:\quad q_0 \rightarrow \frac{1}{q_0} ~.
\end{equation}
By computing the dual period $a^{\text{D}}\equiv a^{\text{D}}_3$ in terms of $a\equiv a_1$, we find
\begin{equation}
a^{\text{D}}= (\ii-1)\tau_1 a
\label{adualSU4}
\end{equation}
in agreement with (\ref{admassless}) for $\omega=\ii$.

\subsection{$N=6$}

We now turn to the last arithmetic case, namely the SU$(6)$ theory. 
We have verified using localization techniques up to two instantons
that in the special vacuum the period matrix can be written as a sum of three independent structures
as follows
\begin{equation}
\Omega = \tau_1\, \mathcal{M}_1+ \tau_2\, \mathcal{M}_2 + \tau_3\, \mathcal{M}_3 ~,
\label{omegaSU6}
\end{equation}
where 
\begin{equation}
\label{Mjforsu6}
\mathcal{M}_1 =
\begin{pmatrix}
\,\frac{1}{3}\,&\,\frac{1}{2}\,&\,\frac{1}{3}\,&\,0\,&\,-\frac{1}{6}\,\\
\,\frac{1}{2}\, &\,1\,&\,1\, &\,\frac{1}{2}\, &\,0\,\\
\,\frac{1}{3}\, &\,1\,&\,\frac{4}{3}\, &\,1\,&\,\frac{1}{3}\, \\
\,0\, &\,\frac{1}{2}\, &\,1\,&\,1\,&\,\frac{1}{2}\, \\
\,-\frac{1}{6}\,&\,0\,&\,\frac{1}{3}\,&\,\frac{1}{2}\,&\,\frac{1}{3}\,
\end{pmatrix},~
\mathcal{M}_2 = 
\begin{pmatrix}
\,1\,&\,\frac{1}{2}\, &\,0\, &\,1\,&\,\frac{1}{2}\, \\
\,\frac{1}{2}\, &\,1\,&\,0\,&\,\frac{1}{2}\, &1\,\\
\,0\,&\,0\,&\,0\,&\,0\,&\,0\,\\
\,1\,&\,\frac{1}{2}\, &\,0\, &\,1\,&\,\frac{1}{2}\, \\
\,\frac{1}{2}\, &\,1\,&\,0\,&\,\frac{1}{2}\, &\,1
\end{pmatrix},~
\mathcal{M}_3=
\begin{pmatrix}
\,\frac{2}{3}\,&\,0\,&\,\frac{2}{3}\,&\,0\,&\,\frac{2}{3}\,\\
\,0\,&\,0\,&\,0\,&\,0\,&\,0\,\\
\,\frac{2}{3}\,&\,0\,&\,\frac{2}{3}\,&\,0\,&\,\frac{2}{3}\,\\
\,0\,&\,0\,&\,0\,&\,0\,&\,0\,\\
\,\frac{2}{3}\,&\,0\,&\,\frac{2}{3}\,&\,0\,&\,\frac{2}{3}\,\\
\end{pmatrix}
\end{equation}
and
\begin{subequations}
\label{uvirmapsu6}
\begin{align}
2\pi\ii\,\tau _1&= \log q_0 + \ii \pi -\log 432+\frac{5}{18}\,q_0 +\frac{485}{5184}\,q_0^2 +\cdots~,
\label{tau1SU6}\\
2\pi\ii\,\tau_2&=\log q_0 + \ii \pi -\log 27 +\frac{4}{9}\,q_0+\frac{14}{81}\,q_0^2 +\cdots~,
\label{tau2SU6}\\
2\pi\ii\,\tau_3 &= \log q_0 + \ii \pi -\log 16 +\frac{1}{2}\,q_0+\frac{13}{64}\,q_0^2 +\cdots~.\label{tau3SU6}
\end{align}
\end{subequations}
We easily recognize that the $q_0$-expansion of $\tau_2$ is the same as that of the effective coupling 
of the SU(3) theory  (see (\ref{q0tausu3})), and that the $q_0$-expansion of $\tau_3$ is the same 
as that of the coupling $\tau_2$ appearing in the SU(4) theory which, as we already remarked, is also the 
same as the coupling $\tau_1$ of the SU(2) theory. Again these facts are a consequence of the symmetries
of the formula (\ref{lambdak}) which imply that these pairs of couplings have the same transformations under
S-duality.

Inverting the expansions (\ref{uvirmapsu6}), we obtain
\begin{subequations}
\label{q0qkSU6}
\begin{align}
q_0 &= - 432 \, q_1 \, (1+ 120\, q_1+4140 \,q_1^2 + \cdots)~,\label{q1SU6}\\
q_0 &= - 27 \, q_2 \, (1+12\, q_2+90\, q_2^2 + \cdots)= 
-27\left(\frac{\eta(3\tau_2)}{\eta(\tau_2)}\right)^{12}~, 
\label{q2SU6}\\
q_0 &= - 16 \, q_3 \, (1+8\,q_3^2+44\,q_3^2 + \cdots)= 
-16\left(\frac{\eta(4\tau_3)}{\eta(\tau_3)}\right)^{8}~,
\label{q3SU6}
\end{align}
\end{subequations}
where we have used the notation (\ref{qk}). We observe that there appears to be no simple way to express 
$q_0$ in terms of $\eta$-quotients of $\tau_1$. 
However, we will revisit this issue in Section~\ref{Saction} where we will provide for all SU$(N)$ models
a universal formula for $q_0$ in terms of modular functions of \emph{any} renormalized couplings $\tau_k$,
thus including also the $\tau_1$ of the SU$(6)$ theory.

Let us now consider the S-duality action on the period matrix (\ref{omegaSU6}).
Under the Sp(10,$\mathbb{Z}$) transformation given in (\ref{STmatrices}), we find that the three matrices 
(\ref{Mjforsu6}) remain invariant while the couplings transform as
\begin{equation}
S\quad:\quad\tau_k \rightarrow-\frac{1}{\lambda_k\,\tau_k}
\end{equation}
with $\lambda_1=1$, $\lambda_2=3$ and $\lambda_3=4$ in full agreement with (\ref{lambdak}).
Exploiting the $\eta$-quotient expressions in \eqref{q0qkSU6}, one can easily prove that the S-transformations 
of $\tau_2$ and $\tau_3$ lead again to 
\begin{equation}
S\quad:\quad q_0 \rightarrow \frac{1}{q_0} ~.
\end{equation}
Finally, by computing the dual period $a^{\text{D}}\equiv a^{\text{D}}_5$ in terms of $a\equiv a_1$ in the
special vacuum, we obtain
\begin{equation}
a^{\text{D}} = -a \tau_1
\end{equation}
which confirms once more (\ref{admassless}), since for SU(6) we have $\omega=\eee^{\pi\ii/3}$.

\vspace{10pt}
Besides the S-duality action we should also consider the T-duality transformation of the effective couplings
which is simply\,\footnote{See Section \ref{SU2special} for the SU$(2)$ case, which has a distinct $T$-
transformation.}
\begin{equation}
T\quad:\quad \tau_k\rightarrow \tau_k+1~.
\end{equation}
Thus the previous results can be summarized by saying that in the arithmetic cases 
the duality transformations act as fractional linear transformations on each of the 
$\tau_k$ and form a subgroup of PSL(2,$\mathbb{R}$) generated by
\begin{equation}
\label{eq:SandTforGammaStar}
S = \begin{pmatrix}
0&1/\sqrt{\lambda_k}\\
-\sqrt{\lambda_k} &0 
\end{pmatrix}\quad\mbox{and}\quad
T = \begin{pmatrix}
~1~&~1~\\
~0~ &~1~ 
\end{pmatrix}
\end{equation}
with $\lambda_k\in\{1,2,3,4\}$ as given by (\ref{lambdak}). We call this subgroup $\Gamma^*(\lambda_k)$.
For $\lambda_k = 1$ this is the usual modular group PSL(2,$\mathbb{Z}$).

\section{\boldmath S-duality and $j$-invariants for the arithmetic theories}
\label{Saction}

In this section we collect the results obtained so far and explain how our definition of S-duality 
fits in the general discussion presented
in \cite{Gaiotto:2009we}. If we describe the SU$(N)$ theory in the special vacuum by
the Seiberg-Witten curve in the Gaiotto form
\begin{equation}
x^N =\frac{u_N}{t^{N-1}(t-1)(t-q_0)}~,
\label{SWG}
\end{equation}
then S-duality can be described as an action on the $(x,t)$ variables given by \cite{Gaiotto:2009we}
\begin{equation}
S\quad:\quad (x\,,\,t) \rightarrow \left(-t^2x\,,\,\frac{1}{t}\right)~,
\label{SdualG}
\end{equation}
which effectively amounts to an inversion of the bare coupling\,\footnote{The curve (\ref{SWG}) retains its 
form if (\ref{Sonq0}) is accompanied by
$u_N\rightarrow(-1)^Nu_N/q_0$.}:
\begin{equation}
\label{Sonq0}
S\quad:\quad\ q_0 \rightarrow \frac{1}{q_0}~.
\end{equation}
We have already seen in various explicit examples that the rule (\ref{Sonq0}) is
implied by the S-duality transformations of the renormalized couplings $\tau_k$ of the arithmetic theories, 
namely
\begin{equation}
\label{Sontauj}
S\quad:\quad \tau_k\rightarrow -\frac{1}{\lambda_k\, \tau_k} 
\end{equation}
with $\lambda_k \in \{2,3,4\}$. Actually, all these cases can be combined together by observing that 
the $\eta$-quotients in (\ref{q0SU2}), (\ref{eta3}), (\ref{q0qkSU4}), (\ref{q0qkSU6}) can be written as
\begin{equation}
\label{q0aseta}
q_0 = - (\lambda_k)^{\frac{6}{\lambda_k-1}}\left(
\frac{\eta(\lambda_k\, \tau_k)}{\eta(\tau_k)} \right)^{\frac{24}{\lambda_k-1}}~,
\end{equation}
from which the inversion rule (\ref{Sonq0}) immediately follows upon using the transformation 
properties of the Dedekind $\eta$-function under (\ref{Sontauj}).
The only case that is not covered by this formula is the relation between $q_0$ and $\tau_1$
in the SU(6) theory, given by the first line of (\ref{q0qkSU6}), for which there seems to be no simple expression 
in terms of $\eta$-quotients.\footnote{Note that in this case we have $\lambda_1=1$ which cannot be used in 
(\ref{q0aseta}).} 
However, the argument based on the transformation properties of the
curve (\ref{SWG}) is completely general; thus, also in this case 
the S-duality transformation $\tau_1\to-1/\tau_1$ should imply, for consistency, an inversion of $q_0$.
We will solve this problem in the following subsections, and in doing so we will actually find a new way of writing 
a non-perturbatively exact relation between the bare coupling and the effective ones. This will turn out to be 
valid not only in all arithmetic cases, including the SU$(6)$ theory mentioned above, 
but also in the non-arithmetic theories, thus opening the way to make further progress. Before
doing this, however, we briefly revisit the SU$(2)$ theory in order to clarify some issues that are  specific to 
the $N=2$ case.

\subsection{The S-duality group of the SU(2) gauge theory}
\label{SU2special}

In the SU$(2)$ gauge theory with four fundamental flavours there is only one renormalized coupling constant 
$\tau_1$, which is related to the bare coupling constant by the non-perturbative relation \eqref{q0aseta} 
with $\lambda_1=4$. This might seem unfamiliar, given that it was already proven in \cite{Seiberg:1994aj} that 
the S-duality group for this theory is the full modular group PSL$(2,\mathbb{Z})$. We now explain how this 
enhancement takes place within the formalism of our paper. 

Let us rewrite the non-perturbative relation between the bare coupling and the renormalized coupling using 
the standard Jacobi $\theta$-functions as follows:
\begin{equation}\label{SU2nonpert}
q_0 = -\left( \frac{\theta_2(2\tau_1)}{\theta_4(2\tau_1)}\right)^4 ~.
\end{equation}
One can check that this coincides with the $\eta$-quotient expression in \eqref{q0SU2}. 
We have already seen that the S-transformation acts as follows on the renormalized coupling $\tau_1$:
\begin{equation}\label{Sontau1su2}
S\quad:\quad \tau_1 \rightarrow -\frac{1}{4\tau_1}~.
\end{equation}
The key point is that only for the SU$(2)$ theory, there is a shift symmetry of the form 
\begin{equation}\label{Tshiftsu2}
T\quad:\quad \tau_1 \rightarrow \tau_1+\frac{1}{2}~.
\end{equation}
This is because, in the presence of massless hypermultiplets in the doublet {pseudoreal} representation, the 
SU(2) theory enjoys a shift symmetry of the effective $\theta$-angle: 
\begin{equation}
\theta\rightarrow \theta +\pi~,
\end{equation} 
which implies \eqref{Tshiftsu2} (see for example appendix B.3 of \cite{Alday:2009aq}). Defining 
$\widetilde{\tau}=2\tau_1$, we see that
(\ref{Sontau1su2}) and the above shift become, respectively, 
$\widetilde{\tau}\rightarrow -{1}/{\widetilde{\tau}}$ and $\widetilde{\tau}\rightarrow \widetilde{\tau}+1$, 
which generate the modular group PSL$(2,\mathbb{Z})$ in full agreement with the original analysis of 
\cite{Seiberg:1994aj}.

Using the non-perturbative relation \eqref{SU2nonpert}, the $T$-transformation (\ref{Tshiftsu2}) 
leads to the following action on the bare coupling constant:
\begin{equation}
T\quad:\quad q_0 ~\rightarrow~ \frac{q_0}{q_0-1} ~.
\end{equation}
Note that this symmetry transformation exists only for the SU$(2)$ gauge theory because in 
all other cases the $T$-action leaves the bare coupling invariant, since it shifts $\tau$ by an integer. 
Combined with the $S$-transformation, which inverts $q_0$, one can check that
\begin{equation}
T S T\quad:\quad q_0 ~\rightarrow~ 1-q_0 ~.
\label{TST}
\end{equation}
 
We now show that this is completely consistent with the Gaiotto formulation of the S-duality group on the bare 
coupling constant. The Gaiotto curve for the SU$(2)$ case is \cite{Gaiotto:2009we}:
\begin{equation}
x^2 =\frac{u_2}{t(t-1)(t-q_0)}~.
\end{equation}
In this expression there is a symmetry between the poles 
at $t=0$ and $t=1$.\footnote{For generic $N$, there is a higher order pole at $t=0$.} Thus, besides 
\eqref{SdualG}, there is another transformation which leaves the curve invariant, namely \cite{Gaiotto:2009we}
\begin{equation}
\widetilde{T}\quad:\quad (x\,,\,t) \rightarrow \left(x\,,\,1-t\right)~.
\label{TdualG}
\end{equation}
It is easy to check that $\widetilde{T}$ precisely generates the transformation \eqref{TST}. Therefore, in the 
SU$(2)$ theory the S-duality group is enhanced to the full modular group PSL(2,$\mathbb{Z}$), on 
account of the half-integer shift of the $\tau$-parameter. 

\subsection{Renormalized couplings and $j$-invariants}

Let us now return to the issue of finding a non-perturbative relation between the renormalized coupling $\tau_1$ 
of the SU$(6)$ theory and the bare coupling constant. The new and key ingredient is the Klein $j$-
invariant function $j(\tau_1)$ for the modular group PSL(2,$\mathbb{Z}$) which is the S-duality group for the
$\tau_1$ coupling of the SU(6) theory. 
The $j$-invariant has the following weak-coupling expansion
\begin{equation}
j(\tau_1) = \frac{1}{q_1}+744+196844\,q_1+21493760\,q_1^2+\cdots~,
\label{jexp}
\end{equation}
with $q_1=\eee^{2\pi\ii\tau_1}$, and is such that 
\begin{equation}
j(\ii)=1728~,\quad j\big(\eee^{\frac{2\pi\ii}{3}}\big)=0\quad\mbox{and}\quad j(\ii\infty)=\infty~.
\label{jvalues}
\end{equation}
The $j$-invariant is also called \textit{hauptmodul} (see for example \cite{GannonHaupt}), 
and is such that all rational functions of $j$ are modular. 

Using (\ref{jexp}), it is possible to verify that
\begin{equation}
\label{mapfortau10}
\frac{\sqrt{j(\tau_1) - 1728} - \sqrt{j(\tau_1)}}{\sqrt{j(\tau_1) - 1728} + \sqrt{j(\tau_1)}} =
- 432 \, q_1 \, (1 + 120 q_1 + 4140 q_1^2 + \cdots)
\end{equation}
which is precisely the same expansion appearing in the first line of (\ref{q0qkSU6}) that was obtained
by inverting the instanton series. Based on this evidence, we propose
that the exact relation between the bare coupling $q_0$ and the renormalized coupling $\tau_1$ is
\begin{equation}\label{mapfortau1}
q_0 = \frac{\sqrt{j(\tau_1) - 1728} - \sqrt{j(\tau_1)}}{\sqrt{j(\tau_1) - 1728} + \sqrt{j(\tau_1)}} ~.
\end{equation}
Further evidence in support of this proposal is its behaviour under $\tau_1\to-1/\tau_1$. This is derived 
from the monodromy of $j$ around the fixed point of this action, {\it{i.e.}} $\tau_1=\ii$, namely
\begin{equation}
\Big(j(\tau_1)-1728\Big)\,\rightarrow\,\eee^{2\pi\ii}\,\Big(j(\tau_1)-1728\Big)~,
\end{equation}
which implies the inversion of $q_0$ as it should be.

This approach is easily generalized, since hauptmoduln have been studied for
the duality groups $\Gamma^*(\lambda_k)$ of the arithmetic theories.\footnote{Recall that the duality group is generated by $S$ and $T$ as defined in \eqref{eq:SandTforGammaStar}.} Indeed, following \cite{Zagier:traces}
for $\lambda_k\in\{1,2,3,4\}$  we introduce the functions $j_{\lambda_k}$ given by\,
\begin{subequations}
\begin{align}
j_1(\tau) &= \left(\frac{E_4(\tau)}{\eta^8(\tau)}\right)^3~,\label{jalpha1}\\
j_2(\tau)&=\left[\left(\frac{\eta(\tau)}{\eta(2\tau)}\right)^{12}
+64\left(\frac{\eta(2\tau)}{\eta(\tau)}\right)^{12}\,\right]^2~,\label{jalpha2}\\
j_3(\tau)&=\left[\left(\frac{\eta(\tau)}{\eta(3\tau)}\right)^{6}
+27\left(\frac{\eta(3\tau)}{\eta(\tau)}\right)^{6}\,\right]^2~,\label{jalpha3}\\
j_4(\tau)&=\left[\left(\frac{\eta(\tau)}{\eta(4\tau)}\right)^{4}
+16\left(\frac{\eta(4\tau)}{\eta(\tau)}\right)^{4}\,\right]^2~.\label{jalpha4}
\end{align}
\label{jalphak}
\end{subequations}
where in the first line $E_4$ is the Eisenstein series of weight 4. It is possible to check that
$j_1$ coincides with the $j$-invariant introduced above, while $j_2$, $j_3$ and $j_4$ are 
generalizations thereof.\footnote{Our definition 
of the $j$-invariants differ from those in \cite{Zagier:traces} by a constant term, which does not affect its 
invariance under the duality group.} Notice that in \eqref{jalpha2}--\eqref{jalpha4} we find 
precisely the $\eta$-quotients appearing in
the relations between the bare coupling $q_0$ and the renormalized couplings $\tau_k$. Solving for these
quotients in terms of the $j_{\lambda_k}$'s and inserting the result in (\ref{q0aseta}), we obtain
\begin{equation}\label{UVIRgeneral}
q_0 = \frac{\sqrt{j_{\lambda_k}(\tau_k) -d_{\lambda_k}^{-1}} 
- \sqrt{ j_{\lambda_k}(\tau_k)}}{\sqrt{j_{\lambda_k}(\tau_k) -d_{\lambda_k}^{-1}} 
+ \sqrt{ j_{\lambda_k}(\tau_k)}} 
\end{equation}
where
\begin{equation}
d_{2}^{-1}=256~,\quad d_{3}^{-1}=108~,\quad d_{4}^{-1}=64~.
\label{dks}
\end{equation}

\begin{figure}
\begin{center}
\includegraphics[scale=1.35]{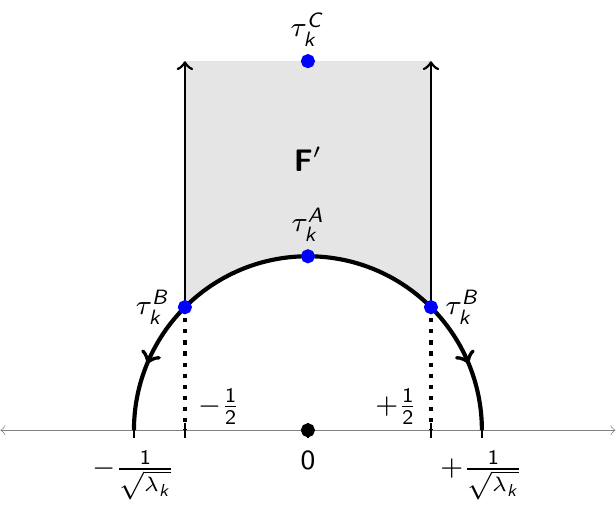}
\caption{The fundamental domain $F'$ of $\Gamma^*(\lambda_k)$. The point $\tau_k^A$ is the fixed point 
of the $S$,
$\tau_k^B$ is the fixed point of $ST^{-1}$, while $\tau_k^C$ is the
fixed point of $T$.}
\end{center}
\end{figure}

Eq.\,(\ref{UVIRgeneral}) has the same structure as (\ref{mapfortau1}); however, this is more than a formal 
analogy. On consulting Fig.~1, one sees that the location of the corners of the fundamental domain --- which are
the fixed points of the $S$, $ST^{-1}$, and $T$ transformations --- are given by
\begin{equation}
\label{cusplocations}
\tau_k^A = \frac{\ii}{\sqrt{\lambda_k}}~ , \quad \tau_k^B = \frac{1}{2} 
+ \frac{\ii }{2}\,\sqrt{\frac{4-\lambda_k}{\lambda_k}}~, \quad \tau_k^C=\ii\infty ~ ,
\end{equation}
respectively. Furthermore, one can show that \cite{Zagier:traces}
\begin{equation}
j_{\lambda_k}(\tau_k^A) = d_{\lambda_k}^{-1}~, 
\quad j_{\lambda_k}(\tau_k^B)=0~,\quad \text{and}\quad j_{\lambda_k}(\tau_k^C)=\infty~,
\label{jatcusp}
\end{equation}
which is a direct generalization of (\ref{jvalues}), while from the monodromy 
of $j_{\lambda_k}$ around the fixed points of $S$, namely
\begin{equation}
\Big(j(\tau_{\lambda_k})-d_{\lambda_k}^{-1}\Big)\,\rightarrow\,\eee^{2\pi\ii}\,
\Big(j(\tau_{\lambda_k})-d_{\lambda_k}^{-1}\Big)~,
\end{equation}
one easily deduces from (\ref{UVIRgeneral}) that $q_0$ gets inverted under S-duality, as expected.

In Tab. 1 we collect the relevant properties of these $j$-invariants together with their expansions around the
cusp point at infinity. In particular we observe in the last column 
that the weak-coupling expansions of the bare coupling $q_0$ are
in perfect agreement with the results presented in Section~\ref{sec:arithmeticity}.
\begin{table}[ht]
\label{UVIRtable}
\begin{center}
\begin{tabular}{|c|c|c|c|}
\hline
$\phantom{\Big|}\lambda_k$ & $d_{\lambda_k}^{-1}$  & $q$-expansion of $j_{\lambda_k}$   & 
$4d_{\lambda_k}\, q_0$\\
\hline\hline
\phantom{\Big|}1 &1728 & $q^{-1}+ 744+196884\,q+21493760\,q^2+\cdots$ &$ q (1 + 120 \,q + 4140 
\,q^2 + \cdots)$\\
\hline
\phantom{\Big|}2 &256 & $q^{-1} + 104+ 4372 \,q + 96256 \,q^2 +\cdots$& $q (1+24 \,q+300 \,q^2 +
\cdots)$\\
\hline
\phantom{\Big|}3 &108 & $q^{-1} +42+ 783 \,q + 8672 \,q^2 +\cdots$ & $q(1+12\,q+90\,q^2+\cdots)$ \\
\hline
\phantom{\Big|}4 &64 & $ q^{-1}+ 24+276\, q + 2048 \,q^2 +\cdots$ & $q(1+8\,q+44\,q^2+\cdots)$ \\
\hline
\end{tabular}
\end{center}
\caption{Relevant parameters for the $j_{\lambda_k}$ functions, their $q$-expansions, and the weak-coupling
expansion of the bare coupling $q_0$ defined in (\ref{UVIRgeneral}).}
\end{table}

\section{\boldmath SU$(N)$ theories and triangle groups}
\label{secn_triangle}

We now proceed to generalize the discussion of the 
previous sections to SU$(N)$ SQCD theories with arbitrary $N$.
To this end, we note that for the arithmetic cases --- $\lambda_k\in\{1,2,3,4\}$ --- the S-duality groups
$\Gamma^*(\lambda_k)$ are particular instances of Hecke groups. A Hecke group $\text{H}(p)$ is a discrete 
subgroup of PSL(2,$\mathbb{R}$) whose generators $T$ and $S$ satisfy
\begin{equation}
S^2=1~,\quad \big(ST\big)^p=1
\label{hecke}
\end{equation}
where $p$ is an integer $\geq 3$.\footnote{The constraints (\ref{hecke}) are usually implemented by
\begin{equation*}
S~:~\quad\widetilde{\tau}~\rightarrow ~-\frac{1}{\widetilde{\tau}}\quad\mbox{and}\quad
T~:~\widetilde{\tau}~\rightarrow~\widetilde{\tau}+2\cos \left(\frac{\pi}{p}\right)~.
\end{equation*} 
By setting $\widetilde{\tau}=2\cos \left(\frac{\pi}{p}\right)\tau$, we see that on $\tau$ 
the group $\text{H}(p)$ coincides with $\Gamma^*(\lambda)$ 
with $\lambda= 4\cos^2 (\frac{\pi}{p})$.} When $p=3$ the Hecke group is the modular group PSL(2,$\mathbb{Z}$).

Using the results of Section~\ref{sec:arithmeticity}, it is not difficult to realize that
$\Gamma^*(\lambda_k)= \text{H}(p_k)$ where
\begin{equation}
\lambda_k = 4\cos^2 \left(\frac{\pi}{p_k}\right)~.
\label{lambdakpk}
\end{equation}
For the four arithmetic cases the correspondence between $\lambda_k$ and $p_k$
is summarized in Tab.~2.
\begin{table}[ht]
\label{lambdap}
\begin{center}
\begin{tabular}{|c||c|c|c|c|}
\hline
$\phantom{\Big|}\lambda_k$ & 1 & 2 & 3 & 4 \\
\hline
$\phantom{\Big|}p_k$ & 3 & 4 & 6 &$\infty$\\ 
\hline
\end{tabular}
\end{center}
\caption{The correspondence between $\lambda_k$ and $p_k$ according to (\ref{lambdakpk}) in the 
arithmetic cases.}
\end{table}

Notice that these are the only cases in which both $\lambda_k$ and $p_k$ are integers. By combining 
(\ref{lambdak}) and (\ref{lambdakpk}), we find
\begin{equation}
\frac{1}{p_k}=\frac{1}{2}-\frac{k}{N}
\label{pk}
\end{equation}
for $k=1,\ldots,\left[\frac{N}{2}\right]$. This formula can be formally extended beyond the arithmetic cases 
where, in general, $p_k$ becomes a rational number.

The Hecke groups $\text{H}(p)$ also exist when $p\not\in\{3,4,6,\infty\}$; moreover they admit a 
generalization into the so-called triangle groups \cite{Doran:2013npa} which we conjecture can 
be further extended for rational $p$. In the following we show that the action of the S-duality group on the
renormalized couplings of the SU($N$) SQCD theories for arbitrary $N$ is precisely 
that of a generalized triangle group.
Furthermore, we show that the $j$-invariant or hauptmodul associated to these triangle groups appears in the 
non-perturbative relation between the bare coupling and the renormalized ones, exactly as in the arithmetic 
cases.

\subsection{A short digression on triangle groups}
\label{subsecn:triangle}
We follow closely the presentation of \cite{Doran:2013npa}, often considering special cases of their formulas 
for our purposes.

Triangle groups are defined by a triple of integer numbers $m_i$ that form the so-called type 
$\mathbf{t} = \left(m_1,m_2,m_3\right)$ and correspond to the orders of the stabilizers. 
These groups are Fuchsian, {\it{i.e.}} they are discrete subgroups  of $\text{PSL}(2,\mathbb{R})$. 
The type $\mathbf{t}$ 
defines a set of angular parameters $v_i = 1/m_i$, which are related to deficit angles $\pi v_i$
at the cusps of the corresponding fundamental domain. In what follows, we will analyze in particular 
types of the form $\mathbf{t} = \left(2,p,\infty\right)$ corresponding to the Hecke groups $\text{H}(p)$
if $p$ is an integer, 
and to their generalizations if $p$ is a rational number. In the latter case the associated triangle
groups are not discrete. 

Let us first consider a type $\mathbf{t} = \left(m_1,m_2,\infty\right)$. Using Theorem 1 
of \cite{Doran:2013npa}, we define the parameter $d_{\mathbf{t}}$ according to
\begin{align}\label{jvalue}
d_{\mathbf{t}}^{-1} =  b' d' &\prod_{k=1}^{b'-1} \left( 2-2 \cos 
\left( 2\pi \frac{k}{b'}\right)\right)^{-\frac{1}{2} \cos \left(\frac{2\pi k a'}{b'}\right)} 
~ \prod_{\ell=1}^{d'-1} 
\left( 2-2 \cos \left(\frac{2\pi\ell}{d'}\right)\right)^{-\frac{1}{2} \cos \left( 2\pi \frac{l c'}{d'}\right)}
\end{align}
where the primed variables are given by
\begin{equation}
\frac{a'}{b'}= \frac{1+v_1-v_2}{2} \quad \text{and} \quad \frac{c'}{d'}= \frac{1+v_1+v_2}{2}
\end{equation}
with $v_i=1/m_i$. Introducing the rescaled variable
\begin{equation}
\widetilde{q} = \frac{q}{d_{\mathbf{t}}} 
\end{equation}
with $q=\eee^{2\pi\ii\tau}$, the hauptmodul $J_\mathbf{t}$
for this triangle group has a weak-coupling expansion
in $\widetilde{q}$ of the form
\begin{equation}
\label{jtansatz}
J_\mathbf{t} \left( \tau \right) =\frac{1}{\widetilde{q}} + \sum_{k=0}^{\infty} c_k \,\widetilde{q}^k \,.
\end{equation}
The coefficients $c_k$ are uniquely determined by the following Schwarzian equation
\begin{equation}\label{jtequation}
-2 \dddot{J_\mathbf{t}} \,\dot{J_\mathbf{t}} + 3 \ddot{J_\mathbf{t}}^2 =
\dot{J_\mathbf{t}}^4 \left(\frac{1-v_2^2}{J_\mathbf{t}^2} + \frac{1-v_1^2}{(J_\mathbf{t}-1)^2}
+\frac{v_1^2+v_2^2-1}{J_\mathbf{t}(J_\mathbf{t}-1)} \right)~.
\end{equation}
Here the dots denote the logarithmic $\tau$-derivatives. The hauptmodul that will be relevant for us is the
one whose weak-coupling expansion begins with $q^{-1}$. This is simply obtained by 
rescaling $J_{\mathbf{t}}$ according to 
\begin{equation}
\label{jresc}
j_{\mathbf{t}}(\tau) = \frac{J_\mathbf{t} (\tau)}{d_{\mathbf{t}}} ~.
\end{equation}

Let us check these formulas for $\mathbf{t}=(2,3,\infty)$ which corresponds to
$\text{H}(3)=\text{PSL}(2,\mathbb{Z}$). When $p=3$ 
the corresponding $\lambda$ is 1 as we see from Tab.~2, and thus instead of the subscript 
${}_{\mathbf{t}}$ we can use the subscript ${}_1$ in all relevant quantities. In this case we have
\begin{equation}
v_1 = \frac{1}{2}\quad\text{and}\quad v_2 = \frac{1}{3}~,
\end{equation}
and
\begin{equation}
\frac{a'}{b'} = \frac{7}{12}\qquad \frac{c'}{d'} = \frac{11}{12} ~.
\end{equation}
Substituting this into (\ref{jvalue}), we find
\begin{equation}
d_1^{-1}={1728}~,
\end{equation}
while the Schwarzian equation (\ref{jtequation}) becomes
\begin{equation}
-2 \dddot{J_1}\, \dot{J_1} + 3 \ddot{J_1}^2 = \dot{J_1}^4 \left(\frac{32-41J_1
+36J_1^{\,2}}{36J_1^{\,2}(J_1^{\,2}-1)} \right)~.
\end{equation}
Solving for $J_1$ and rescaling the solution with $d_1$ according to (\ref{jresc}), one gets
\begin{equation}
j_1(\tau)= 1728\, J_1(\tau)= \frac{1}{q} + 744+ 196884\,q+ \cdots 
\end{equation}
which exactly matches the expansion of the absolute $j$-invariant of the modular group (see (\ref{jexp})).

In a similar way one can check that for $\mathbf{t}=(2,p,\infty)$
with $p\in\{4,6,\infty\}$, the above formulas correctly lead to the expressions of the $j$-invariants
and the $d$ parameters of the other arithmetic cases that are summarized in Tab.~1. However, as we have 
already mentioned, these same formulas can be used also for other integer values of $p$ and formally
extended to the
case in which $p$ is a rational number. As a first example of this extension we consider the 
SU(5) SQCD theory.

\subsection{$N=5$}
Using localization techniques we have computed the prepotential and the period matrix of the SU(5) theory with 
10 massless flavours up to 2 instantons. In the special vacuum we find that the period matrix 
$\Omega$ can be conveniently written as a sum of two independent structures, in agreement with the general 
formula (\ref{Omegafull}). Defining
\begin{equation}
\begin{aligned}
\lambda_1&\,=\, 4\sin^2\frac{\pi}{5}\,=\,4\cos^2\frac{3\pi}{10}\,=\,\frac{\sqrt{5}}{2}
\left(\sqrt{5}-1\right)~,
\\
\lambda_2&\,=\, 4\sin^2\frac{2\pi}{5}\,=\,4\cos^2\frac{\pi}{10}\,=\,\frac{\sqrt{5}}{2}\left(\sqrt{5}+1\right)~,
\end{aligned}
\label{lambda12}
\end{equation} 
the quantum corrected period matrix can be written as
\begin{equation}
\Omega = \tau_1 \, {\mathcal M}_1 + \tau_2\, {\mathcal M}_2
\end{equation}
where
\begin{equation}
{\mathcal M}_1 = \begin{pmatrix}
\frac{2 \lambda_1}{5} & \frac{1}{\lambda_1} &\frac{\lambda_1}{5} & -\frac{\lambda_1^2}{5 \sqrt{5}}\\
\frac{1}{\lambda_1} & \frac{2}{\lambda_1} & \frac{\sqrt{5}}{\lambda_1^2} & \frac{\lambda_1}{5} \\
\frac{\lambda_1}{5}  & \frac{\sqrt{5}}{\lambda_1^2} & \frac{2}{\lambda_1} &\frac{1}{\lambda_1} \\
-\frac{\lambda_1^2}{5 \sqrt{5}} &\frac{\lambda_1}{5} &\frac{1}{\lambda_1} &\frac{2 \lambda_1}{5}
\end{pmatrix}~,\qquad
{\mathcal M}_2 = \begin{pmatrix}
\frac{2}{\lambda_1} & \frac{\lambda_1}{5} & \frac{1}{\lambda_1} & \frac{\sqrt{5}}{\lambda_1^2} \\
\frac{\lambda_1}{5} &\frac{2 \lambda_1}{5} & -\frac{\lambda_1^2}{5 \sqrt{5}}&\frac{1}{\lambda_1}\\
\frac{1}{\lambda_1} & -\frac{\lambda_1^2}{5 \sqrt{5}}& \frac{2 \lambda_1}{5}&\frac{\lambda_1}{5} \\
\frac{\sqrt{5}}{\lambda_1^2}&\frac{1}{\lambda_1}&\frac{\lambda_1}{5}&\frac{2}{\lambda_1}
\end{pmatrix}~,
\end{equation}
and
\begin{align}\label{uvirtau1}
2\pi \ii \, \tau_1 &= \log q_0 + \ii \pi - \log \left[25 \sqrt{5}\left(\frac{2}{\sqrt{5}-1}\right)^{\sqrt{5}}\,
\right]
+\frac{8 q_0}{25}+\frac{14 q_0^2}{125} + \cdots ~, \\
2 \pi \ii \, \tau_2 &= \log q_0 + \ii \pi - \log \left[25 \sqrt{5}\left(\frac{2}{\sqrt{5}+1}\right)^{\sqrt{5}}\,
\right]+\frac{12 q_0}{25}+\frac{24 q_0^2}{125} + \cdots \label{uvirtau2}~.
\end{align}
This structure is less cumbersome than it appears at first sight. Indeed, one can check that the parameters
$\lambda_1$ and $\lambda_2$ in (\ref{lambda12}) are another instance of the general formula 
(\ref{lambdak}) and that 
\begin{equation}
\mathcal{M}_1+\mathcal{M}_2 = \mathcal{C}
\end{equation}
where $\mathcal{C}$ is the Cartan matrix of SU(5). Moreover, the classical and the logarithmic terms 
of $\tau_1$ and $\tau_2$ exactly coincide with the results already reported in \cite{Minahan:1996ws}. 
But, most importantly, using the S-duality transformations described in Section~\ref{Sduality}, 
one finds that the matrices $\mathcal{M}_1$
and $\mathcal{M}_2$ remain invariant while the effective couplings transform simply as
\begin{align}
\tau_1 \rightarrow -\frac{1}{\lambda_1\, \tau_1}\quad\text{and}\quad \tau_2\rightarrow 
-\frac{1}{\lambda_2\, \tau_2} ~.
\end{align}
These observations show that the SU(5) theory has the same general features we encountered 
in the arithmetic cases. Therefore it is natural to expect that also this theory can be understood 
along the same lines, and in particular that it is possible to write
non-perturbatively exact expressions for the relations between the bare coupling and 
the renormalized ones in terms of hauptmoduln. We now confirm that these expectations are correct.

Let us first put $k=2$. The form of $\lambda_2$ in (\ref{lambda12}) indicates to us that the relevant
Hecke group is $\text{H}(10)$. Indeed, for $k=2$ and $N=5$, equation (\ref{pk}) yields $p_2=10$
so that the type of the triangle group is $\mathbf{t}_2=(2,10,\infty)$. Using this in (\ref{jvalue}), 
with a little bit of algebra we obtain
\begin{equation}
d_{\lambda_2}^{-1} = 4\left[25 \sqrt{5}\left(\frac{2}{\sqrt{5}+1}\right)^{\sqrt{5}}\,\right]=76.2385\cdots 
~,
\label{d2}
\end{equation}
while from (\ref{jtequation}) we get the following rescaled hauptmodul
\begin{equation}
j_{\lambda_2}(\tau_2) = \frac{1}{q_2} + \frac{19}{50}\frac{1}{ d_{\lambda_2}} + \frac{673}{10000}
 \frac{ q_2}{d_{\lambda_2}^2} + \frac{701 }{93750} \frac{q_2^2}{ d_{\lambda_2}^3} + \cdots
\end{equation}
with $q_2=\eee^{2\pi\ii\tau_2}$. This function is such that 
\begin{equation}
j_{\lambda_2}(\tau_2^A) = d_{\lambda_2}^{-1}~, 
\quad j_{\lambda_2}(\tau_2^B)=0\quad \text{and}\quad j_{\lambda_2}(\tau_2^C)=\infty
\label{jatcusp2}
\end{equation}
where $\tau_2^{A,B,C}$ are the cusp locations in the $\tau_2$-plane given by (\ref{cusplocations}) with 
the current value of $\lambda_2$. Notice also that the quantity in square brackets in (\ref{d2}) also appears
in the 1-loop logarithmic term of (\ref{uvirtau2}). 

These facts and our experience with the arithmetic theories indicate that it is in fact not too bold to propose 
that the relation between the bare coupling $q_0$ and the renormalized coupling $\tau_2$ be of the general form 
(\ref{UVIRgeneral}), namely
\begin{equation}
q_0 = \frac{ \sqrt{j_{\lambda_2}(\tau_2) - d_{\lambda_2}^{-1}} 
- \sqrt{ j_{\lambda_2}(\tau_2)} }{ \sqrt{j_{\lambda_2}(\tau_2) - d_{\lambda_2}^{-1}} 
+ \sqrt{ j_{\lambda_2}(\tau_2)} } 
= - \frac{q_2}{4 d_{\lambda_2}}\left( 1+\frac{3}{25}\frac{q_2}{d_{\lambda_2}}
+\frac{6}{625}\frac{q_2^2}{d_{\lambda_2}^2} + \cdots\right)~.
\end{equation}
Inverting this series and taking the logarithm, we obtain
\begin{equation}
2\pi\ii\,\tau_2 = \log q_0 + \ii \pi +\log\big(4d_{\lambda_2}\big) 
+\frac{12}{25}q_0+\frac{25}{125}q_0^2 + \cdots
\end{equation}
which precisely matches the instanton expansion for $\tau_2$ in \eqref{uvirtau2} obtained 
using equivariant localization! Furthermore, from the monodromy around $\tau_2^A$ which
is the fixed point under $S$, namely
\begin{equation}
\Big(j(\tau_{\lambda_2})-d_{\lambda_2}^{-1}\Big)\,\rightarrow\,\eee^{2\pi\ii}\,
\Big(j(\tau_{\lambda_2})-d_{\lambda_2}^{-1}\Big)~,
\end{equation}
we see that $q_0$ gets inverted, in agreement with the general expectations.
This analysis shows that the action of the S-duality group on the effective coupling $\tau_2$ of the SU(5) 
theory is that of the Hecke group H$(10)$.

We now turn to $k=1$ and the quantum coupling $\tau_1$. The form of $\lambda_1$ in (\ref{lambda12})
indicates that we are dealing with a non-Hecke group. Indeed, setting $k=1$ and $N=5$ in (\ref{pk}), we get 
$p_1=\frac{10}{3}$ which leads to the type $\mathbf{t}_1=(2,\frac{10}{3},\infty)$. Despite the
non-integer entry of $\mathbf{t}_1$, we still proceed and apply the formulas we have 
described in the previous subsection to obtain $d_{\lambda_1}$ and the hauptmodul 
$j_{\lambda_1}(\tau_1)$. Specifically, from (\ref{jvalue}) after some algebraic manipulations we get
\begin{equation}
d_{\lambda_1}^{-1}=4\left[25 \sqrt{5}\left(\frac{2}{\sqrt{5}-1}\right)^{\sqrt{5}}\,\right]= 655.8364\cdots~,
\end{equation}
while from the Schwarzian equation (\ref{jtequation}) we find the following rescaled hauptmodul
\begin{equation}
j_{\lambda_1}(\tau_1) = \frac{1}{q_1} + \frac{21}{50}\frac{1}{d_{\lambda_1}} + \frac{663}{10000} 
\frac{q_1}{d_{\lambda_1}^2}+\frac{227}{46875} \frac{q_1^2}{d_{\lambda_1}^3} + \cdots
\end{equation}
with $q_1=\eee^{2\pi\ii\tau_1}$. This function is such that
\begin{equation}
j_{\lambda_1}(\tau_1^A) = d_{\lambda_1}^{-1}~, 
\quad j_{\lambda_1}(\tau_1^B)=0\quad \text{and}\quad j_{\lambda_1}(\tau_1^C)=\infty
\label{jatcusp1}
\end{equation}
where $\tau_1^{A,B,C}$ are the three cusps in the $\tau_1$-plane (see (\ref{cusplocations})). 
Plugging these results into our universal formula (\ref{UVIRgeneral}), we get
\begin{equation}
q_0 =\frac{ \sqrt{j_{\lambda_1}(\tau_1) - d_{\lambda_1}^{-1}} 
- \sqrt{ j_{\lambda_1}(\tau_1)} }{ \sqrt{j_{\lambda_1}(\tau_1) - d_{\lambda_1}^{-1}} 
+ \sqrt{ j_{\lambda_1}(\tau_1)} } 
= - \frac{q_1}{4 d_{\lambda_1}}\left( 1+\frac{2}{25}\frac{q_1}{d_{\lambda_1}}+
\frac{13}{5000}\frac{q_1^2}{d_{\lambda_1}^2} + \cdots\right)~.
\end{equation}
Inverting this and taking the logarithm of $q_1$, we obtain
\begin{equation}
2\pi\ii\,\tau_1 = \log q_0+ \ii \pi +\log\big(4d_{\lambda_1}\big)
 +\frac{8}{25}q_0+\frac{14}{125}q_0^2 + \cdots
\end{equation}
which is in perfect agreement with the explicit result (\ref{uvirtau1}) derived from localization!
Again, from the monodromy around $\tau_1^A$, which is the fixed point of $S$, 
we easily see that under S-duality $q_0$ is correctly mapped into its inverse.

In conclusion, the SU$(5)$ theory has two non-arithmetic couplings which are related to the bare coupling
by the same universal formula that holds in the arithmetic theories.

\subsection{Generalization to higher $N$}
The analysis of the previous subsection can be extended to arbitrary values of $N$. Even if the algebraic
manipulations become more and more involved as $N$ increases, it is possible to prove that the quantum period 
matrix can always be written as
\begin{equation}
\label{omegadecompagain}
\Omega = \sum_{k=1}^{\big[\frac{N}{2}\big]} \tau_k \, {\mathcal{M}}_k 
\end{equation}
where each individual coefficient $\tau_k$ transforms under the duality group according to
\begin{equation}
S \quad : \quad \tau_k \rightarrow - \frac{1}{\lambda_k \,\tau_k}\qquad\mbox{and}
\qquad 
T\quad:\quad \tau_k \rightarrow \tau_k+1
\label{STagain}
\end{equation}
for some positive $\lambda_k$.

To show this, let us first consider $N$ to be an odd number. In this case a careful analysis 
\cite{Minahan:1997fi} of the duality transformations on the homology cycles of the Seiberg-Witten curve shows 
that
$S$ and $T$ have to satisfy the constraint
\begin{equation}
\left(S\,T\,S^{-1}\,T\right)^N= \mathbf{1}~.
\label{STconstraint}
\end{equation}
 Given (\ref{STagain}), it is not difficult to show that
 \begin{equation}
S\,T\,S^{-1}\,T=\begin{pmatrix}
1&1\\
-\lambda_k &1-\lambda_k 
\end{pmatrix}~.
\end{equation}
 The $N^{\text{th}}$ power of this matrix projectively equals the identity, as required by (\ref{STconstraint}), 
 if
 \begin{equation}
 \lambda_k = 4 \sin^2 \left( \frac{k\, \pi}{N} \right)\,=\,4\cos^2\left(\frac{(N-2k)\,\pi}{2N}\right)
 \label{lambdakagain}
 \end{equation}
 or
 \begin{equation}
 \lambda_k = 4 \cos^2 \left( \frac{k\, \pi}{N} \right)~.
 \label{lambdakcos}
 \end{equation}
The latter solution, however, leads to an additional constraint of the form $(S\,T)^N=-1$, which is not found 
in the explicit realization of the $S$ and $T$ transformations as Sp$(2N-2,\mathbb{Z})$ matrices 
\cite{Minahan:1997fi}. This leaves us with the solution (\ref{lambdakagain}) which is precisely the spectrum we 
conjectured and found to be true in all cases we have considered so far. 

The matrices $\mathcal{M}_k$ can be given an explicit expression too. The key ingredient for this is the matrix 
$\mathcal{G}$ appearing at 1-loop (see (\ref{G})). Decomposing it into its $\left[\frac{N}{2}\right]$ 
independent components according to
\begin{equation}
\mathcal{G}= \sum_{k=1}^{\big[\frac{N}{2}\big]} \log\sin\left(\frac{k\,\pi}{N}\right)\,\mathcal{G}_k~,
\label{Gk}
\end{equation}
it turns out that the matrices
\begin{equation}
\label{Mk}
\mathcal{M}_k= \sum_{\ell=1}^{\big[\frac{N}{2}\big]} \lambda_{k\ell}\,\,\mathcal{G}_\ell ~=~ 4 
\sum_{\ell=1}^{\big[\frac{N}{2}\big]} \sin^2\left(\frac{k\,\ell\,\pi}{N}\right)\,\mathcal{G}_\ell 
\end{equation}
satisfy the required properties. Notice also that these matrices add up to the Cartan matrix:
\begin{equation}
\sum_{k=1}^{\big[\frac{N}{2}\big]}\mathcal{M}_k = \mathcal{C}~.
\end{equation}
We have explicitly checked and verified these statements up to $N=15$.

Having the spectrum of the allowed $\lambda_k$'s, from the cosine expression in (\ref{lambdakagain}) we see 
that the type of the generalized triangle group that we should consider is
\begin{equation}
\mathbf{t}_k=\Big(2,\frac{2N}{N-2k},\infty\Big)
\label{typek}
\end{equation}
whose second entry is in general a rational number. As we have seen in  the SU(5) theory, there are no 
obstructions in extending the formulas (\ref{jvalue}), (\ref{jtansatz}) and (\ref{jresc}) to
types with a rational entry. Thus, proceeding as we described in the previous subsections, we can determine 
$d_{\mathrm{t}_k}$ and the hauptmoduln $j_{\mathrm{t}_k}$ corresponding to (\ref{typek}) 
and use the resulting expressions into the universal formula (\ref{UVIRgeneral}) to find the exact relation between 
the bare coupling $q_0$ and the renormalized one $\tau_k$. If this procedure is correct, 
inverting this map order by order in $q_0$ we should retrieve the multi-instanton expansion produced by the 
localization method, exactly as we showed for $N=5$. 
In Appendix~\ref{secn:N7} we give some details for the case $N=7$, where again we finding perfect agreement. 
At this point, it should be clear that our procedure works for arbitrary values of $N$.
We regard the complete agreement between these two approaches as a highly non-trivial and quite remarkable 
check on the consistency of the procedure.

The above results are valid also when $N$ is even. In this case, the spectrum of $\lambda_k$ is still 
given by (\ref{lambdakagain}) while the matrices $\mathcal{G}_{k}$ and $\mathcal{M}_k$ are defined by 
(\ref{Gk}) and (\ref{Mk}) with the caveat that for $k=\frac{N}{2}$ one should use
\begin{equation}
\mathcal{G}_{\frac{N}{2}}=\, \mathcal{C}-\sum_{k=1}^{\frac{N}{2}-1}\mathcal{G}_k\qquad\mbox{and}
\qquad \mathcal{M}_{\frac{N}{2}}= \,\mathcal{C}-\sum_{k=1}^{\frac{N}{2}-1}\mathcal{M}_k~.
\label{GMn2}
\end{equation}
We have checked this is indeed the case up to $N=14$. 

\subsection{Relation to earlier work}

We now show that our analysis is consistent with earlier discussions of S-duality in conformal SQCD theories and 
that it extends them  in several aspects. Consider the Seiberg-Witten curve (\ref{SW}) for the massless case 
and in the special vacuum: 
\begin{equation}
y^2 = (x^N-u_N)^2 - h\, x^{2N} ~.
\end{equation}
Using our results, we can write the function $h$ in terms of the renormalized couplings as follows:
\begin{equation}
h=\frac{4q_0}{(1+q_0)^2}=\frac{1}{1-d_{\lambda_k}\, j_{\lambda_k}(\tau_k)}~.
\label{hofj}
\end{equation}
This shows that for any $N$ the Seiberg-Witten curve can be expressed in terms of $j$-invariants. Of course, 
any of the renormalized couplings can be chosen as long as the appropriate 
$j$-invariant is used. 

Let us now consider the behaviour of $h$ near the cusp points (\ref{cusplocations}). Using (\ref{jatcusp}), it is 
easy to find that 
\begin{equation}
\begin{aligned}
h(\tau_k) &\rightarrow \infty \quad \ \text{near}\ \tau_k^A ~,\\
h(\tau_k) &\rightarrow 1 \quad \ ~\,\text{near}\ \tau_k^B ~,\\
h(\tau_k) &\rightarrow 0 \quad \ ~\,\text{near}\ \tau_k^C ~.
\end{aligned}
\label{hbehavior}
\end{equation}
Given the meaning of the fixed points, we conclude that the monodromy of $h$ around $\infty$, $1$ and $0$
yields, respectively, the behaviour under $S$, $ST^{-1}$ and $T$. This is precisely what we began with in 
Section~\ref{Sduality}. There, we 
obtained the $T$ and $S$ matrices by associating them with monodromies around the points $h=0$ 
and $h=\infty$, respectively, and by following their effects on the $\hat{\alpha}$- and $\hat{\beta}$-cycles 
of the Seiberg-Witten curve. It is reassuring to rederive this very same result by studying the 
action of the duality group on the quantum couplings $\tau_k$. This provides additional confirmation 
for our proposal (\ref{UVIRgeneral}).

There are a number of novel elements in our discussion compared with earlier 
works \cite{ Minahan:1996ws,Argyres:1998bn,Minahan:1997fi}. To begin with, we note that
(\ref{UVIRgeneral}) 
represents a non-perturbatively exact relation between the bare and renormalized coupling constants. As we 
have shown in a case-by-case study, this completely specifies the manner in which all $\left[ \frac{N}{2} \right]$ coupling constants are renormalized for all SU$(N)$ theories in the special vacuum.

Furthermore, we observe that previous investigations have focused on a specific renormalized coupling, which in 
our notation, is $\tau_{\left[\frac{N}{2}\right]}$. For odd $N$, the type (\ref{typek}) corresponding to
$k=\big[\frac{N}{2}\big]$ is $(2, 2N, \infty)$, which identifies the Hecke group H$(2N)$, while for even $N$, 
the type becomes $(2,\infty,\infty)$ corresponding to the Hecke group H($\infty$) which is isomorphic 
to $\widetilde{\Gamma}^0 (2)$. Thus, we have successfully reproduced the observations 
of \cite{Argyres:1995wt,Argyres:1997cg,Minahan:1997fi} that these Hecke groups are relevant when 
considering the duality properties of SU($N$) theories. However, as we have tried to emphasize, 
one does not need to single out any specific
quantum coupling $\tau_k$ in order to understand the S-duality group. Indeed, one could 
choose to express $q_0$ in terms of any of the $\tau_k$'s since the behaviour of the curve near the cusps is 
universal and independent of this choice. 

While this remains true away from the massless limit (provided the mass deformations are turned on in a 
controlled manner), we find that expressing the observables in terms of specific effective coupling 
constants $\tau_k$ instead of the bare coupling $q_0$ expedites the identification of modular structures. In 
particular, the choice of which effective couplings to consider follows solely from S-duality 
constraints. This, in turn, makes it possible to resum the non-perturbative data of the gauge theory into modular 
forms associated to congruence subgroups of the full modular group. This analysis is the subject of Part II.

\part{Massive $\mathcal N=2$ SQCD and modular anomaly equations}
In this part, we discuss $\mathcal{N}=2$ SQCD theories with $2N$ massive fundamental hypermultiplets in
the special vacuum.
In order to retain the $\mathbb{Z}_N$ symmetry of the special vacuum, we will consider only
mass configurations that preserve this symmetry. 
Furthermore, we will restrict our attention to the arithmetic theories.
The reason for this is just a matter of simplicity. Indeed, as we will see, in the arithmetic theories 
the S-duality groups $\Gamma^*(\lambda_k)$ contain subgroups that are also congruence 
subgroups of the modular group PSL(2,$\mathbb{Z}$), so that the analysis of the modular 
properties of the various observables can be done using standard modular forms, without the need of 
introducing the more involved theory of automorphic forms.
Since the SU$(2)$ and SU$(3)$ SQCD theories have already been considered from this 
point of view in \cite{Billo:2013fi,Billo:2013jba} and in \cite{Ashok:2015cba} respectively, 
we will discuss in detail the other
two arithmetic cases, namely $N=4$ and $N=6$, even if many of the subsequent formulas 
are valid
for arbitrary $N$.

\section{Mass deformations and observables}
\label{secn:masses}
While the classical prepotential (\ref{Fclass}) is unaffected by mass deformations, the 
1-loop prepotential (\ref{F1loop}) becomes
\begin{equation}
F_{\text{$1$-loop}} = \sum_{u \neq v=1}^N \gamma(A_u - A_v) 
- \sum_{u=1}^{N}\sum_{f=1}^{2N} \,\gamma(A_u +m_f)~.
\label{F1loopm}
\end{equation}
Expanding for small masses, one obtains an expression in which the $2N$ fundamental masses appear through
the Casimir invariants of the flavour group, namely
\begin{equation}
T_\ell = \sum_{f=1}^{2N} \big(m_{f}\big)^{\ell}
\label{massinv}
\end{equation}
for $\ell=1,\ldots,2N$. As we mentioned above, in order not to spoil the $\mathbb{Z}_N$ 
symmetry of the special vacuum, we turn on only those flavour Casimirs that are $\mathbb{Z}_N$-symmetric. 
This can be done by choosing the following mass configuration
\begin{align}\label{specialmasses}
m_f = \begin{cases}
\omega^{f-1} \ m &, \quad f \in \left\lbrace 1, \cdots, N \right\rbrace ~,\\
\omega^{f-1}\, \widetilde{m} &, \quad f \in \left\lbrace N+1, \cdots, 2N \right\rbrace~,
\end{cases}
\end{align}
where $\omega=\eee^{\frac{2\pi\ii}{N}}$,  
which in turn implies
\begin{equation}
T_N = N\left(m^N + \widetilde{m}^N\right)\quad\text{and}\quad 
T_{2N} = N\left(m^{2N} + \widetilde{m}^{2N}\right) \, ,
\label{TN2N}
\end{equation}
with all other $T_\ell$ vanishing. In what follows, by special vacuum we will mean 
both the restriction (\ref{specialvacuum}) on the scalar vacuum expectation values and the above choice of 
masses.

As discussed in Section \ref{subsecn:observ}, the $\mathbb{Z}_N$-invariance of the special vacuum
implies that the prepotential has a semi-classical expansion of the form (\ref{Fsv}), but now the coefficients
$f_n$ depend also on the mass invariants (\ref{TN2N}), namely
\begin{equation}
F = \sum_{n} \frac{f_n\left(q_0; T_N, T_{2N}\right)}{a^{Nn}}~.
\end{equation}
The $f_n$'s must have mass-dimension equal to $(nN+2)$, but since $q_0$ is dimensionless and 
$T_N$ and $T_{2N}$ have dimensions $N$ and $2N$ respectively, it is not possible to satisfy 
this requirement. As a result, in the massive case as well, the special vacuum prepotential vanishes identically.

Let us now turn to the dual period $a^{\text{D}}$. When the 1-loop and instanton corrections are
taken into account, we find
\begin{equation}
a^{\text{D}} = c_N\, a\, \tau_1 +\frac{c_N}{2\pi\ii}\,
\sum_{n=0}^\infty\frac{g_n^{(1)}(\tau_1;T_N,T_{2N})}{a^{Nn+N-1}}
\label{adualmass}
\end{equation}
where $c_N$ is defined in (\ref{cNgeneral}).
This form, which will be confirmed by the explicit examples worked out in the later sections, 
can be argued simply using dimensional analysis because $g_n^{(1)}$ has mass 
dimension $(Nn+N)$ and can be constructed out of the $\mathbb{Z}_N$-invariant Casimirs $T_N$ 
and $T_{2N}$.

Finally, we consider the period matrix $\Omega$. Its decomposition in terms of the matrices $\mathcal{M}_k$ 
that diagonalize the $S$-action remains valid
\begin{equation}
\label{Omegadecom}
\Omega = \widetilde{\tau}_1\, \mathcal{M}_1 + \widetilde{\tau}_2 \,\mathcal{M}_2 + \cdots  ~,
\end{equation}
but now the coefficients acquire terms proportional to the flavour Casimirs. In particular one finds
\begin{equation}\label{tautildeexpmass}
\widetilde{\tau}_k = \tau_k - \frac{1}{2\pi \ii} \sum_{n=0}^{\infty}(Nn+N-1)\,
\frac{g_{n}^{(k)}(\tau_k; T_N, T_{2N})}{a^{Nn+N}}
\end{equation}
for $k=1,\cdots,\left[\frac{N}{2}\right]$. Detailed examples will be given in the following sections.

\section{S-duality in massive SQCD}
\label{secn:sdualitymass}

To see the implications of S-duality in massive SQCD theories, we use the same approach described in 
\cite{Ashok:2015cba} for the SU$(3)$ theory and introduce the following combination
\begin{equation}
\begin{aligned}
\label{Xexp}
X &:= a^{\text{D}} - c_N\, a\, \tau_1 \\
&= \frac{c_N}{2\pi \ii}\sum_n \frac{g_n}{a^{Nn+N-1}}
\end{aligned}
\end{equation}
where $g_n\equiv g_n^{(1)}(\tau_1; T_N, T_{2N})$.
We now perform an S-duality transformation on the first line of (\ref{Xexp}) and use (\ref{saadual}), 
(\ref{stau}) and (\ref{lambda10});
after some simple algebra we get 
\begin{equation}
S(X)= \frac{1}{c_N\,\omega^2\,\tau_1}\,X~.
\end{equation}
On the other hand, applying S-duality to the second line of (\ref{Xexp}) we get
\begin{equation}
S(X) =\frac{ c_N}{2\pi \ii}\sum_n \frac{S(g_n)}{(-\omega \,a^{\text{D}})^{Nn + N-1}}~.
\end{equation}
If we now substitute the expression (\ref{adualmass}) for $a^{\text{D}}$ and equate the two different 
expressions for $S(X)$ order by order in the large-$a$ expansion, we can deduce how the coefficients
$g_n$ transform under $S$. From the leading term, we simply find
\begin{equation}\label{Song0}
S(g_0) = \left(\ii\sqrt{\lambda_1}\tau_1\right)^{N-2}g_0~,
\end{equation}
where $\lambda_1$ is as in (\ref{lambda10}). For the higher order terms, however, we find non-linear 
contributions that lead to a recursion relation
\begin{equation}
S(g_n)=(-1)^{n} \left(\ii\sqrt{\lambda_1}\tau_1\right)^{Nn+N-2}\, \left(g_n + \frac{1}{2\pi \ii\tau_1}
\sum_m (Nm+N-1)g_m g_{n-m-1} +\cdots\,\right)~.
\end{equation}
The summand on the right hand side is symmetric under $m\rightarrow (n-m-1)$, and thus $S(g_n)$ can be
more conveniently written as
\begin{equation}
\label{gnunderS}
S(g_n) =(-1)^{n} \left(\ii\sqrt{\lambda_1}\tau_1\right)^{Nn+N-2}
 \left(g_n + \frac{(Nn+N-2)}{4\pi \ii\tau_1}\sum_m g_m \,g_{n-m-1} +\cdots\,\right)~.
\end{equation}
The presence of the $(-1)^n$ factor suggest to us that the notion of $S$-parity or charge under S-duality will 
be a useful one. We define it to be $(+1)$ when $n$ is even and $(-1)$ when $n$ is odd.

So far $N$ has been generic, but to make further progress from now on we will
restrict our attention to the arithmetic cases for which $\lambda_1$ is an integer. In fact, in these cases the S-
duality group $\Gamma^*(\lambda_1)$ contains a subgroup, denoted as $\Gamma_1(\lambda_1)$, 
which is also a congruence subgroup of PSL(2,$\mathbb{Z}$). 
The modular forms of such a subgroup, which are
well-known and classified (see for instance \cite{Koblitz,Apostol}), will play a crucial role in our analysis and will 
appear in the exact expressions of the coefficients $g_n$. To see this, let us first recall that
$\Gamma_1(\lambda_1)$ is generated by $T$ and $S'=STS^{-1}$, the latter acting on the effective
coupling as
\begin{equation}
S'\quad:\quad \tau_1 \rightarrow \frac{\tau_1}{1-\lambda_1 \tau_1}~.
\end{equation}
When $\lambda_1$ is an integer, this is indeed an element of PSL(2,$\mathbb{Z})$. 
Combining the actions of $S$ and $T$, we can easily deduce how the conjugate 
periods $a$ and $a^{\text{D}}$ transform under $S'$. The result is 
\begin{equation}
S'(a^{\text{D}}) = a^{\text{D}} \quad \mbox{and}\quad
S'(a) = a + \omega(1-\omega)a^{\text{D}} ~. 
\end{equation}
Using these rules on $X$, from the first line of (\ref{Xexp}) we get
\begin{equation}
S'(X) = \frac{1}{1-\lambda_1\tau_1}\, X ~,
\end{equation}
while from the second line of (\ref{Xexp}) we find
\begin{equation}
S'(X) = \frac{c_N}{2\pi \ii}\sum_n \frac{S'(g_n)}{((1-\lambda_1\tau_1) a)^{Nn+N-1}}\left(1+
\frac{c_N}{2\pi \ii (1-\lambda_1 \tau_1)}\sum_m\frac{g_m}{a^{Nm+N-2}}\right)^{Nn+N-1}~.
\end{equation}
Equating these two expressions, to leading order we obtain
\begin{equation}\label{S'ong0}
S'(g_0) = (1-\lambda_1\tau_1)^{N-2}\, g_0~,
\end{equation}
while at higher orders we get a recursion relation very similar to the one obtained before for $S$, namely
\begin{equation}\label{gnunderS'}
S'(g_n) = (1-\lambda_1 \tau_1)^{Nn+N-2} \left(g_n + \frac{(Nn+N-2)}{4\pi \ii\tau_1}
\sum_m g_m \,g_{n-m-1} +\cdots\right)~.
\end{equation}
Eq. (\ref{S'ong0}) shows that $g_0$ is a modular form of $\Gamma_1(\lambda_1)$ with weight $(N-2)$.
As we will see in the specific examples in the next section, such a modular form behaves under $S$ 
exactly as required by (\ref{Song0}), thus proving the consistency of our analysis. On the other hand,
the presence of non-linear terms in the right hand side of (\ref{gnunderS'}) implies that 
the coefficients $g_n$ for $n>0$ are \emph{quasi}-modular forms of $\Gamma_1(\lambda_1)$ with 
weight $(Nn+N-2)$ that satisfy a modular anomaly equation to which we now turn.

\subsection{The modular anomaly equation}
In \cite{Billo:2013fi,Billo:2013jba,Ashok:2015cba} it has been shown that in the massive
SU$(2)$ and SU$(3)$ theories the quasi-modularity is due to the presence of the anomalous
Eisenstein series $E_2$. 
The same conclusion has been reached for the $\mathcal{N}=2^\star$ theories with arbitrary gauge groups
in \cite{Billo:2014bja,Billo':2015ria,Billo':2015jta}. Therefore it is very natural to expect that for
the massive higher rank SQCD theories too, the Eisenstein series $E_2$ plays a fundamental role.

Let us recall that $E_2$ is a quasi-modular form of weight 2 such that
\begin{equation}
E_2\Big(\!\!-\frac{1}{\tau_1} \Big) =- \big(\ii\tau_1\big)^2\, \Big(E_2(\tau_1)+
\frac{6}{\ii \pi\tau_1}\Big)~.
\label{E2transf}
\end{equation}
In the arithmetic cases under consideration, it is always possible to form a linear combination 
of $E_2$ and a modular form of $\Gamma_1(\lambda_1)$, which under the $S$ transformation
$\tau_1\to-\frac{1}{\lambda_1\tau_1}$ transforms in a way similar to (\ref{E2transf}). More precisely,
if we denote such a combination by $\widetilde{E}_2^{(\lambda_1)}$, we will have
\begin{equation}
\widetilde{E}_2^{(\lambda_1)} \left( -\frac{1}{\lambda_1 \tau_1} \right) 
= -\left( \ii \sqrt{\lambda_1} \tau_1 \right)^2 
\left( \widetilde{E}_2^{(\lambda_1)}(\tau_1) + \frac{6}{\ii \pi \tau_1} \right) ~.
\label{SE2tilde}
\end{equation}
Notice that the existence of such a combination is a priori not obvious 
since the $S$-transformation lies outside both the modular group and its 
congruence subgroup $\Gamma_1(\lambda_1)$. 
Nevertheless this combination exists and the explicit examples for the relevant cases are given
in Appendix~\ref{secn:gamma} (see in particular (\ref{E22tilde}), (\ref{E23tilde}) and (\ref{E24tilde})). 

Following \cite{Billo:2013fi,Billo:2013jba,Ashok:2015cba} we propose that the coefficients $g_n$ 
depend on $\tau_1$ only through $\widetilde{E}_2^{(\lambda_1)}$ and the modular forms of
$\Gamma_1(\lambda_1)$, in such a way that they are globally quasi-modular forms of 
$\Gamma_1(\lambda_1)$ with total weight $(Nn+N-2)$. For simplicity, in the following we will only exhibit the 
dependence on $\widetilde{E}_2^{(\lambda_1)}$ and just write
$g_n\big[\widetilde{E}^{(\lambda_1)}_2\big]$. Then, applying S-duality, we have
\begin{equation}
\begin{aligned}
S\left(g_n\big[\widetilde{E}^{(\lambda_1)}_2\big]\right) 
&= (-1)^n(\ii\sqrt{\lambda_1}\tau_1)^{Nn+N-2}\, g_n\Big[\widetilde{E}_2^{(\lambda_1)} 
+ \frac{6}{\ii \pi \tau_1} \Big]\\
&= (-1)^n(\ii\sqrt{\lambda_1}\tau_1)^{Nn+N-2}\, \left(g_n\big[\widetilde{E}^{(\lambda_1)}_2\big]+
\frac{6}{\ii \pi \tau_1}\frac{\partial g_n}{\partial \widetilde{E}_2^{(\lambda_1)}}+\cdots \right)
\end{aligned}
\end{equation}
where the second line follows upon expanding for large $\tau_1$. Comparing with (\ref{gnunderS}) 
we obtain the modular anomaly equation
\begin{equation}
\label{recursiongeneralN}
\frac{\p g_n}{\p \widetilde{E}_2^{(\lambda_1)}} 
= \frac{Nn+N-2}{24} \sum_{m=0}^{n-1} g_m \,g_{n-m-1} 
\end{equation}
which has the form of a recursion relation.
Indeed, given the initial condition that specifies $g_0$ as
a modular form, the $\widetilde{E}_2$-dependent part of $g_1$ can be unambiguously obtained by integrating 
the modular anomaly equation. This leaves room for a truly modular piece, which can be fixed
by comparing with the explicit instanton expansion obtained using localization. Once $g_1$ is fully fixed, we
can use it in (\ref{recursiongeneralN}) to find $g_2$, and recursively proceed in this way for the higher
$g_n$'s.
This approach has been successfully applied to the SU$(3)$ theory in \cite{Ashok:2015cba}. In the 
next sections we complete the analysis for the SU$(4)$ and SU$(6)$ theories.

\subsection{Coupled modular anomaly equations}
We now consider the period matrix $\Omega$. As we mentioned in Section~\ref{secn:masses}, after including
the quantum corrections it can be decomposed as in (\ref{Omegadecom}) where, under S-duality, the flavour 
deformed couplings $\widetilde{\tau_k}$ transform as
\begin{equation}\label{Sontautilde}
S\quad:\quad \widetilde{\tau}_k \rightarrow -\frac{1}{\lambda_k \widetilde{\tau}_k} ~.
\end{equation}
The fact that $\widetilde{\tau_k}$ behave like $\tau_k$ is a simple consequence of the algebraic 
properties of the matrices $\mathcal{M}_k$. Applying S-duality to both sides of (\ref{tautildeexpmass}), 
we get
\begin{equation}\label{tautildesemiclass}
-\frac{1}{\lambda_k\tau_k}\!\left(\!1-\frac{1}{2\pi \ii \tau_k} \sum_{m=1}^{\infty} \frac{Nm+N-1}{a^{Nm}} 
g_{m-1}^{(k)} \right)^{\!-1}\!\!\!=
-\frac{1}{\lambda_k\tau_k} -\frac{1}{2\pi\ii} \sum_{n=1}^{\infty} \frac{Nn+N-1}{(-\omega 
a^{\text{D}})^{Nn}} S\big(g^{(k)}_{n-1}\big)
\end{equation}
which, after inserting the semi-classical expansion (\ref{adualmass}) for the dual period, yields the S-duality
transformation rules for the coefficients $g_n^{(k)}$. In particular, at leading order we find
\begin{equation}
\label{Song0k}
S \big(g_{0}^{(k)}\big)
= \frac{ \left( \ii \sqrt{\lambda_1} \tau_1 \right)^{N}}{\left( \ii \sqrt{\lambda_k} \tau_k \right)^2} 
\, g_{0}^{(k)}~,
\end{equation}
while at higher orders we get
\begin{equation}
\begin{aligned}
S\big(g_{n}^{(k)}\big) &= \frac{\left(-1\right)^n \left( \ii \sqrt{\lambda_1} \tau_1 \right)^{Nn+N}}{\left( \ii 
\sqrt{\lambda_k} \tau_k \right)^2} \Bigg[ \,g_{n}^{(k)} \\
&\qquad\quad
+ \frac{1}{2 \pi \ii \tau_k} \sum_m \left( \frac{(Nm+N-1)(N(n-m)-1)}{Nn+N-1} g_{m}^{(k)} 
\,g_{n-m-1}^{(k)}\right)\\
&\qquad\quad
+ \frac{1}{2 \pi \ii \tau_1} \sum_m \left( \frac{(Nm+N)(Nm+N-1)}{Nn+N-1} g_{m}^{(k)} \,g_{n-m-1} \right)
+\cdots\Bigg]~.
\end{aligned}
\label{Songnj}
\end{equation}
When $k=1$, both (\ref{Song0k}) and (\ref{Songnj}) reduce to (\ref{Song0}) and (\ref{gnunderS}), 
respectively. This is a simple but important consistency check of our analysis. 

We now perform a similar analysis for the $S'$ transformation under which each 
effective coupling $\widetilde{\tau}_k$ changes as
\begin{equation}
\widetilde\tau_k \rightarrow \frac{\widetilde\tau_k}{1-\lambda_k\, \widetilde\tau_k} ~.
\end{equation}
Since in the arithmetic theories the $\lambda_k$'s are integers, this 
is a PSL(2,$\mathbb{Z}$) transformation.
Using the general technique of comparing coefficients in the semi-classical expansions, we obtain 
the following constraint for $g_0^{(k)}$:
\begin{equation}
\label{S'ong0k}
S'\big(g_{0}^{(k)} \big)
= \frac{ \left( 1-\lambda_1 \tau_1\right)^{N}}{\left( 1-\lambda_k\tau_k \right)^2} \, g_{0}^{(k)}~,
\end{equation}
while for the higher coefficients $g_n^{(k)}$ we get
\begin{equation}
\begin{aligned}
S'\!\big( g_{n}^{(k)} \big)& = \frac{( 1-\lambda_1 \tau_1 )^{Nn+N}}{(1-\lambda_k\tau_k )^2} 
\Bigg[ g_{n}^{(k)} \\
&\qquad\quad+ \frac{1}{2 \pi \ii \tau_k} \sum_m \left( \frac{(Nm+N-1)(N(n-m)-1)}{Nn+N-1} g_{m}^{(k)} 
\,g_{n-m-1}^{(k)}\right) \\
&\qquad\quad
+ \frac{1}{2 \pi \ii \tau_1} \sum_m \left( \frac{(Nm+N)(Nm+N-1)}{Nn+N-1} g_{m}^{(k)} \,g_{n-m-1}
 \right)+\cdots\Bigg]~.
\end{aligned}
\label{S'ongnk}
\end{equation}
Again it is not difficult to check that for $k=1$ these two equations reduce respectively to (\ref{S'ong0}) and 
(\ref{gnunderS'}), as it should be.

{From} (\ref{S'ong0k}) combined with (\ref{Song0k}), we can infer that $g_{0}^{(k)}$ is a ratio of a modular
form of $\Gamma_1(\lambda_1)$ with weight $N$ and a modular form of $\Gamma_1(\lambda_k)$ with 
weight 2. Likewise, by combining (\ref{S'ongnk}) with (\ref{Songnj}) we deduce that for $n>0$ the coefficients
$g_n^{(k)}$ are quasi-modular meromorphic forms of $\Gamma_1(\lambda_1)$ and 
$\Gamma_1(\lambda_k)$ which receive contributions from 
both $\widetilde{E}_2^{(\lambda_1)}$ and $\widetilde{E}_2^{(\lambda_k)}$.
Taking into account the factors multiplying the square brackets in (\ref{Songnj}) and (\ref{S'ongnk}), 
we are led to the following ansatz:
\begin{equation} 
\label{ansatz}
g_n^{(k)} = \sum_{\ell=0}^n G_{n}^{\,nN+N-2\ell; 2+2n-2\ell}(\tau_1,\tau_k)\,
\left(\widetilde{E}_2^{(\lambda_1)}(\tau_1)\right)^{\ell}
\left(\widetilde{E}_2^{(\lambda_k)}(\tau_k)\right)^{n-\ell}
\end{equation}
where the coefficients $G_n^{r_1;r_k}(\tau_1,\tau_k)$ are made of modular forms of
$\Gamma_1(\lambda_1)$ and $\Gamma_1(\lambda_k)$ with weights $r_1$ and $r_k$
respectively. Using the anomalous transformation properties of the second Eisenstein series, from (\ref{ansatz})
we get
\begin{equation}
S\big(g_n^{(k)}\big) = \frac{\left(-1\right)^n \left( \ii \sqrt{\lambda_1} \tau_1 \right)^{Nn+N}}{\left( \ii 
\sqrt{\lambda_k} \tau_k \right)^2} \left( g_{n}^{(k)} + \frac{6}{\pi\ii\tau_1}\frac{\p g_n^{(k)}}{\p 
\widetilde{E}_2^{(\lambda_1)}}  + \frac{6}{\pi\ii\tau_k}\frac{\p g_n^{(k)}}{\p 
\widetilde{E}_2^{(\lambda_k)}}  +\cdots \right)~,
\end{equation}
and, after comparison with (\ref{Songnj}), we arrive at the following coupled equations
\begin{equation}
\begin{aligned}
\frac{\p g_n^{(k)}}{\p \widetilde{E}_2^{(\lambda_k)}} &=\frac{1}{12}\sum_{m=0}^{n-1}\frac{(Nm+N-1)
(N(n-m)-1)}{Nn+N-1} g_{m}^{(k)} \,g_{n-m-1}^{(k)}~,\\
\frac{\p g_n^{(k)}}{\p \widetilde{E}_2^{(\lambda_1)}} &=\frac{1}{12}\sum_{m=0}^{n-1}\frac{(Nm+N)(Nm
+N-1)}{Nn+N-1} g_{m}^{(k)}\, g_{n-m-1}~.
\end{aligned}
\label{coupledeqs}
\end{equation}
In order for these equations to be consistent and integrable, it is necessary that the mixed second derivatives 
computed from either line of (\ref{coupledeqs}) match. We find that this is indeed the case, since we have
\begin{equation}
\frac{\p}{\p \widetilde{E}_2^{(\lambda_1)}}\Bigg(\frac{\p g_n^{(k)} }{\p \widetilde{E}_2^{(\lambda_k)}}
\Bigg)-
\frac{\p}{\p \widetilde{E}_2^{(\lambda_k)}}\Bigg(\frac{\p g_n^{(k)} }{\p \widetilde{E}_2^{(\lambda_1)}}
\Bigg)=0~.
\end{equation}
Given the structure of the modular anomaly equations (\ref{coupledeqs}), this is a non-trivial check which 
makes it possible to ``integrate-in'' the quasi-modular terms in a consistent manner.

\section{\boldmath Resummation: the cases $N=4$ and $N=6$}
\label{secn:SU46}

In this section we study in detail the SU$(4)$ and SU$(6)$ gauge theories along the lines discussed before. 
Throughout this section, we use special cases of the formulas derived in the previous section, i.e.~setting 
$N=4$ or $N=6$ as the case may be.

\subsection{$N=4$}
For the SU$(4)$ theory the relevant parameters are:
\begin{equation}
\omega= \ii~,\qquad c_4=\ii-1~,\qquad k=1,2~,\qquad \lambda_1=2~,\qquad \lambda_2=4~.
\label{su4values}
\end{equation}

\subsubsection{The dual period}
We have computed the SU$(4)$ prepotential, the dual periods, and the period matrix up to three instantons 
using localization methods. 
From these results, after using the relation (\ref{q0qkSU41}) to rewrite the instanton counting parameter $q_0$ 
in terms of the renormalized coupling $q_1$, we find that the dual period can be written as
\begin{equation}
a^\text{D} = ( \ii-1) \, a \, \tau_1  + \frac{(\ii-1)}{2\pi \ii} \sum_{n=0}^{\infty} 
\frac{g_n (q_1; T_4, T_8)}{a^{4n+3}}
\end{equation}
in agreement with the general form (\ref{adualmass}). The first coefficients $g_n$ are
\begin{subequations}
\label{gnsu4}
\begin{align}
g_0 &= \frac{T_4}{12} \left(1+24\,q_1+24 \,q_1^2+96\, q_1^3+ \cdots \right) ~, \label{g0su4}\\
g_1 &= \frac{T_4^2}{4} \left(q_1+26 \,q_1^2+84 \,q_1^3+\cdots\right) 
+\frac{T_8}{56} \left(1-56\, q_1-2296\, q_1^2-13664\, q_1^3+\cdots\right) ~,\label{g1su4} 
\end{align}
\end{subequations}
where, as usual, we have set $q_1=\eee^{2\pi\ii\tau_1}$. 

Our goal is to show that these expressions arise from a weak-coupling expansion of quasi-modular forms
of $\Gamma_1(2)$. Indeed, according to the discussion of the previous section, we should have
\begin{equation}
\label{sgnSU4}
\begin{aligned}
S(g_n)&=(-1)^{n} \big(\ii\sqrt{2}\tau_1\big)^{4n+2}\, \big[\, g_n +\cdots\,\big]\phantom{\Big|}~,\\
S'(g_n) &= (1-2\tau_1)^{4n+2} \,\big[ \,g_n  +\cdots\big]\phantom{\Big|}~.
\end{aligned}
\end{equation}
In particular for $n=0$ when there are no extra terms beyond leading order, these equations tell us that $g_0$ 
should be a modular form of $\Gamma_1(2)$ with weight 2 and $S$-parity $(+1)$. As shown in 
Appendix~\ref{secn:gamma} there is only one such form, namely $f_{2,+}^{(2)}$ whose weak-coupling 
expansion is
\begin{equation}
f_{2,+}^{(2)}=1+24 q_1+24 q_1^2+96 q_1^3+24 q_1^4+144q_1^5 \cdots ~.
\end{equation}
Comparing with (\ref{g0su4}), we are led to conclude
\begin{equation}
g_0 = \frac{T_4}{12} \, f_{2,+}^{(2)}~,
\end{equation}
which, to be consistent with (\ref{sgnSU4}), implies also that $T_4$ is invariant under both $S$ and $S'$ 
transformations, namely $S(T_4)= S'(T_4)=T_4$. 
We would like to stress that once we assume that $g_0$ is
a modular form $\Gamma_1(2)$ of weight 2, the only freedom we have is the overall coefficient which is fixed 
by matching with the perturbative contribution. After this is done, \emph{all} non-perturbative terms are fixed 
by the Fourier expansion of the modular form. The fact that these terms perfectly match the
explicit multi-instanton results coming from localization up to three instantons is a very strong and highly 
non-trivial test of our general strategy. 

To obtain the coefficients $g_n$ for $n>0$ we can use the recursion relation (\ref{recursiongeneralN}), which 
in the present case is
\begin{equation}
\label{eq:recursionsu4}
\frac{\p g_n}{\p\widetilde{E}_2^{(2)}} = \frac{2n+1}{12}\,\sum_{m=0}^{n-1} g_{m}\, g_{n-m-1}
\end{equation}
where $\widetilde{E}_2^{(2)}$ is the quasi-modular form introduced in Appendix~\ref{secn:gamma}
(see in particular (\ref{E22tilde})). 
Let us now determine $g_1$ which according to our general analysis should be
a quasi-modular form of of $\Gamma_1(2)$ with weight 6 and with $S$-parity $(-1)$ that solves
the above modular anomaly equation for $n=1$, namely
\begin{equation}
\frac{\p g_1}{\p\widetilde{E}_2^{(2)}}= \frac{1}{4}\,g_0^2~.
\end{equation}
Integrating with respect to $\widetilde{E}_2^{(2)}$ and using the exact expression for $g_0$ obtained above, 
we find
\begin{equation}
g_1 = \frac{T_4^2}{576} \big(f_{2,+}^{(2)}\big)^2\widetilde{E}_2^{(2)} + \text{modular piece} \, ,
\end{equation}
where by `modular piece' we mean a modular form of $\Gamma_1(2)$ with weight 6 and with $S$-parity 
$(-1)$. As shown in Appendix~\ref{secn:gamma} there is only one such form, namely
\begin{equation}
f_{2,+}^{(2)}\,f_{4,-}^{(2)}= 1 - 56 q_1 - 2296 q_1^2 - 13664 q_1^3+\cdots~.
\end{equation}
Comparing with the localization result (\ref{g1su4}), obtain the following exact expression
\begin{equation}
g_1 = \frac{T_4^2}{576}\Bigg[ \big(f_{2,+}^{(2)}\big)^2\widetilde{E}_2^{(2)}  - \frac{3}{2}\,f_{2,+}^{(2)}
\,f_{4,-}^{(2)}\Bigg] +\, \frac{T_8}{56}f_{2,+}^{(2)}\,f_{4,-}^{(2)}~.
\end{equation}
As before, all coefficients are fixed by matching with the perturbative terms and following that, all non-
perturbative contributions follow from the Fourier expansions of the modular forms. 
The agreement with the explicit multi-instanton results in (\ref{g1su4}) is rather remarkable.

The above procedure can be iteratively used to determine the higher coefficients $g_n$. In this way we
have determined up to $g_3$, always finding perfect agreement with the localization results.

\subsubsection{The period matrix}
In the special vacuum the period matrix $\Omega$ of the massive SU(4) theory
can be compactly written as
\begin{equation}
\Omega =  \widetilde\tau_1\, \mathcal{M}_1 +\widetilde\tau_2\, \mathcal{M}_2 
\end{equation}
where the two matrices $\mathcal{M}_k$ are given in (\ref{M2M4matrices}) and
\begin{align}
\widetilde\tau_k &= \tau_k -\frac{1}{2\pi\ii } \sum_{n=0}^{\infty}\,
\frac{4n+3}{a^{4n+4}} \, \widehat{g}^{(k)}_n \left(q_0; T_4, T_8 \right)~.
\end{align}
This has the same form as (\ref{tautildeexpmass}), except that the coefficients are expressed in terms
of the bare coupling $q_0$ instead of the renormalized ones; this is the meaning of the $\widehat{g}_k^{(n)}$ 
notation.
{From} our explicit calculations, using the non-perturbative relation (\ref{q0qkSU41}) we find 
\begin{equation}
\widehat{g}_n^{(1)}\left(q_0; T_4, T_8 \right) = g_n\left(q_1; T_4, T_8 \right) 
\end{equation}
where the $g_n$'s are the same coefficients appearing in the dual period, for which we have already given 
exact expressions. On the other hand, we find that the first $\widehat{g}_n^{(2)}$ coefficients are
\begin{subequations}
\label{g4expansions}
\begin{align}
\widehat{g}^{(2)}_0 &= \frac{T_4}{12}\, \Big(1 - \frac{1}{4}\,q_0 - \frac{25}{256}\, q_0^2 -
\frac{29}{512}\, q_0^3 + \cdots \Big)~, \label{g02SU4}\\
\widehat{g}^{(2)}_1 &= -\frac{T_4^2}{224}\,\Big( q_0 + \frac{7}{64}\,q_0^2 + \frac{7}{512}\, q_0^3 + 
\cdots \Big)
+ \frac{T_8}{56} \Big(1 - q_0 - \frac{5}{128}\,q_0^2 - \frac{39}{512}\,q_0^3 + \cdots \Big) ~.
\label{g12SU4}
\end{align}
\end{subequations}
The challenge is now to show that, once the bare coupling is mapped into the renormalized ones, the resulting 
expressions $g_n^{(2)}$ have good modular properties. In particular for $g_0^{(2)}$, according to the 
general analysis of the previous section (see (\ref{Song0k}) and (\ref{S'ong0k}) for $N=4$ and $k=2$), we 
should have
\begin{equation}
\label{SDualOnGPlusI}
\begin{aligned}
S \big(g_{0}^{(2)}\big) = \frac{ \left( \sqrt{2}\,\ii\, \tau_1 \right)^{4}}{\big(2 \,\ii\,\tau_2 \big)^2} 
\, g_{0}^{(2)}\qquad\mbox{and}\qquad
S' \big(g_{0}^{(2)}\big) =\frac{ \big(1-2\tau_1 \big)^{4}}{\big(1-4\tau_2 \big)^2} 
\, g_{0}^{(2)}~.
\end{aligned}
\end{equation}
These equations tell us that $g^{(2)}_0$ is the ratio of a modular form of $\Gamma_1 (2)$ in $\tau_1$
with weight 4 and a modular form of $\Gamma_1 (4)$ in $\tau_2$ with weight 2, with total S-parity $(+1)$. 
{From} the list of the modular forms presented in Appendices~\ref{secn:gamma}
for $\Gamma_1(2)$ and $\Gamma_1(4)$, we see
that the most general ansatz which satisfies these properties is
\begin{equation}
g_0^{(2)} = \frac{T_4}{12} \,\Bigg[\,x\, \frac{\big(f_{2,+}^{(2)}\big)^2}{f_{2,+}^{(4)}} 
+\big(1-x\big)\,\frac{f_{4,-}^{(2)}}{f_{2,-}^{(4)}}\,\Bigg]~,
\end{equation}
where the overall coefficient has been fixed to match with the perturbative result in (\ref{g02SU4}) and $x$
is a free parameter. By Fourier expanding the modular forms and expressing the result
in terms of the bare coupling $q_0$, one sees that both meromorphic forms within square brackets are identical
and both match the $q_0$ expansion in (\ref{g02SU4}).
In the following we choose for simplicity $x=1$, so that\,\footnote{We could just as well have picked $x=0$; 
the Fourier expansions do not distinguish between these choices, and it is clear that the modular anomaly 
equations are not affected by this choice.}
\begin{equation}
g_0^{(2)} = \frac{T_4}{12} \,\frac{\big(f_{2,+}^{(2)}\big)^2}{f_{2,+}^{(4)}}~.
\end{equation}
For the higher coefficients $g_n^{(2)}$, we have to use the coupled modular anomaly equations
(\ref{coupledeqs}). For $n=1$ they become
\begin{equation}
\begin{aligned}
\frac{\p g_1^{(2)}}{\p \widetilde{E}_2^{(4)}} = \frac{3}{28}\, \big(g_0^{(2)}\big)^2 \qquad
\mbox{and}\qquad
\frac{\p g_1^{(2)}}{\p \widetilde{E}_2^{(2)}} = \frac{1}{7}\,g_0^{(2)}\,g_0 ~, 
\end{aligned}
\label{modeqsg12}
\end{equation}
where $\widetilde{E}_2^{(2)}$ and $\widetilde{E}_2^{(4)}$ are 
the quasi-modular forms of $\Gamma_1(2)$ and $\Gamma_1(4)$ defined in (\ref{E22tilde}) and
(\ref{E24tilde}) respectively.
Integrating (\ref{modeqsg12}) we find
\begin{equation}
g_1^{(2)} = \frac{3}{28}\, \big(g_0^{(2)}\big)^2 
\widetilde{E}_2^{(4)} +
\frac{1}{7}\, g_0^{(2)}\,g_0 \,\widetilde{E}_2^{(2)} + \text{modular piece} ~.
\end{equation}
As before, the `modular piece' is determined by considerations of weight and S-parity and by demanding 
agreement with the perturbative terms in (\ref{g12SU4}). Explicitly, we find
\begin{equation}
g_1^{(2)} = \frac{T_4^2}{1344} \left( \frac{\big(f_{2,+}^{(2)}\big)^4
\widetilde{E}_2^{(4)}}{\big(f_{2,+}^{(4)}\big)^2} 
+\frac{4}{3}\frac{\big( f_{2,+}^{(2)} \big)^3\widetilde{E}_2^{(2)}}{f_{2,+}^{(4)}}  
-\frac{9}{2}\frac{\big(f_{2,+}^{(2)}\big)^2 f_{4,-}^{(2)}}{f_{2,+}^{(4)}} \right) + \frac{T_8}{56}\, 
\frac{\big(f_{2,+}^{(2)}\big)^2 f_{4,-}^{(2)}}{f_{2,+}^{(4)}}~.
\end{equation}
Once again, the perturbative terms are enough to fix all coefficients that are not determined
by the modular anomaly equations; then, the instanton contributions follow by Fourier expanding the modular
forms. The perfect agreement with the explicit result (\ref{g12SU4}) obtained from localization confirms in
a very non-trivial way the validity of our procedure. 

Using this approach iteratively, we have computed higher $g_k^{(2)}$ coefficients, 
finding complete agreement with the multi-instanton results.

\subsection{$N=6$}
We now repeat the above analysis for the massive SU$(6$) theory. In this case
the relevant parameters are:
\begin{equation}
\omega= \eee^{\frac{\pi\ii}{3}}~,\qquad c_6=-1~,\qquad k=1,2,3~,\qquad \lambda_1=1~,\qquad 
\lambda_2=3~,\qquad \lambda_3=4~.
\label{su6values}
\end{equation}

\subsubsection{The dual period}
The large-$a$ expansion of dual period of the massive SU(6) theory takes the form
\begin{equation}\label{dualperiodmassive}
a^{\text{D}}=-a\, \tau_1 -\frac{1}{2\pi \ii}\sum_n \frac{g_n(q_1; T_6, T_{12})}{a^{6n+5}}~.
\end{equation}
Using localization methods we have computed the coefficients $g_n$ up to two instantons
and rewritten them in terms of the effective parameter $q_1$ by means the relation (\ref{mapfortau1}). 
The explicit expressions of the
first coefficients are
\begin{subequations}
\label{gnlocalization}
\begin{align}
g_0 &= \frac{T_6}{5}\left(1+240q_1+2160q_1^2+\cdots\right)~,
\label{g0su6}\\
g_1&=12\,T_6^2\left(
q_1+258q_1^2+\cdots\right)+\frac{T_{12}}{22}\left(1-264q_1-135432q_1^2+\cdots\right)~.
\label{g1su6}
\end{align}
\end{subequations}
Since $\lambda_1=1$ we expect to resum these expansions into standard modular forms of 
PSL$(2,\mathbb{Z})$. In particular, from (\ref{Song0}) and (\ref{S'ong0}) we have
\begin{equation}
\label{Song0SU40}
S(g_0) = \big(\ii\,\tau_1\big)^4g_0\qquad\mbox{and}\qquad
S'(g_0) = \big(1-\tau_1\big)^4 g_0~,
\end{equation} 
which tell us that $g_0$ is a modular form of weight 4 with positive $S$-parity. 
The unique form of this kind is the Eisenstein series
$E_4$; thus, matching the perturbative contribution we find
\begin{equation}
g_0=\frac{T_6}{5}\, E_4~.
\end{equation}
Again, all instanton terms are dictated by the Fourier expansion of $E_4$ and are in perfect agreement
with the localization result (\ref{g0su6}). We also verify that $g_0$ satisfies (\ref{Song0SU40}) provided
that $T_6$ is invariant under S-duality, namely $S(T_6)=S'(T_6)=T_6$.

To obtain the coefficients $g_n$ with $n>0$, we use the modular anomaly 
equation (\ref{recursiongeneralN}) which in this
case becomes
\begin{equation}
\frac{\partial g_n}{\partial E_2} = \frac{3n+2}{12}\sum_{m=0}^{n-1}g_{m}\,g_{n-m-1} ~.
\end{equation}
For example, integrating this equation for $n=1$ and fixing the $E_2$-independent part by comparing
with the perturbative contributions, we get
\begin{equation}
g_1 = \frac{T_6^2}{60}\,\big(E_4^2\,E_2-E_4\,E_6\big)+\frac{T_{12}}{22}\, E_4\,E_6~.
\end{equation}
Using the Fourier expansion of the Eisenstein series it is easy to check that the instanton terms precisely
match those in (\ref{g1su6}). Proceeding iteratively in this manner one can derive the exact expressions of
the higher coefficients $g_n$. In particular we have explicitly computed a few higher $g_n$, 
always finding perfect agreement with the localization results.

\subsubsection{The period matrix}

In the special vacuum the period matrix $\Omega$ of the massive SU$(6)$ theory can be written compactly as
\begin{equation}
\Omega = \widetilde{\tau}_1\, \mathcal{M}_1+\widetilde{\tau}_2\, \mathcal{M}_2+\widetilde{\tau}_3\, 
\mathcal{M}_3
\end{equation}
where the three matrices $\mathcal{M}_k$ are given in (\ref{Mjforsu6}) while, using a notation similar 
to that of the SU$(4)$ theory, the three effective couplings turn out to have 
the following semi-classical expansion
\begin{equation}
\widetilde{\tau}_k = \tau_k -\frac{1}{2\pi\ii } \sum_{n=0}^{\infty}\,
\frac{6n+5}{a^{6n+6}} \, \widehat{g}^{(k)}_n(q_0; T_6, T_{12}) \, .
\end{equation}
The coefficients $\widehat{g}^{(1)}_n$
coincide with the $g_n$'s already discussed, while the first coefficients for $k=2$ are
\begin{subequations}
\label{g3exps}
\begin{align}
\widehat{g}_0^{(2)} &=\frac{T_6}{5}\,\Big(1- \frac{7}{18}\,q_0 - \frac{319}{2592}\, q_0^2 + \cdots 
\Big)~,
\label{g02SU6}\\
\widehat{g}_1^{(2)}&=-\frac{7\,T_6^2}{198}\Big(q_0 - \frac{65}{504}\,q_0^2 
+\cdots\Big)+\frac{T_{12}}{22}\Big(1 + \frac{7}{9}\, q_0 - \frac{443}{1296}\, q_0^2 +\cdots\Big)~,
\label{g12SU6}
\end{align}
\end{subequations}
and for $k=3$ are
\begin{subequations}
\label{g3exps1}
\begin{align}
\widehat{g}_0^{(3)} &= \frac{T_6}{5}\Big( 1- \frac{1}{3}\, q_0 - \frac{47}{432}\, q_0^2 
+\cdots\Big)~,
\label{g03SU6}\\
\widehat{g}_1^{(3)}&= -\frac{5\,T_6^2}{132}\Big( q_0 - \frac{23}{360}\, q_0^2 +\cdots\Big) +
\frac{T_{12}}{22}\Big(1 + \frac{5}{6}\, q_0 - \frac{227}{864}\,q_0^2 + \cdots \Big)~.
\label{g13SU6}
\end{align}
\end{subequations}
We now show that these are the first few terms in the semi-classical 
expansion of rational functions of quasi-modular forms. Since the procedure is similar 
to that of the SU$(4)$ theory, we will be brief in our discussion.

Let us first consider $g_0^{(2)}$ and $g_0^{(3)}$, whose $S$ and $S'$ transformations are
\begin{equation}
\begin{aligned}
S \big(g_{0}^{(2)}\big)&= \frac{ \left(\ii\, \tau_1 \right)^{6}}{\big(\sqrt{3}\,\ii\,\tau_2 \big)^2} 
\, g_{0}^{(2)}~,\qquad
S' \big(g_{0}^{(2)}\big)= \frac{ \left(1-\tau_1 \right)^{6}}{\big(1-3\tau_2 \big)^2} 
\, g_{0}^{(2)}~,\\
S \big(g_{0}^{(3)}\big)&= \frac{ \left(\ii\, \tau_1 \right)^{6}}{\big(2\,\ii\,\tau_3 \big)^2} 
\, g_{0}^{(3)}~,
\qquad~~\,S' \big(g_{0}^{(3)}\big)= \frac{ \left(1-\tau_1 \right)^{6}}{\big(1-4\tau_3 \big)^2} 
\, g_{0}^{(3)}~.
\end{aligned}
\end{equation}
These formulas suggest that $g_0^{(2)}$ should be expressed as a ratio of a modular form 
in $\tau_1$ with weight 6 and a modular form of $\Gamma_1(3)$ in $\tau_2$ 
with weight 2, with an overall S-parity equal to $(+1)$. Likewise, $g_0^{(3)}$ should be 
expressed as a ratio of a modular form in $\tau_1$ with weight 6 and a modular
form of $\Gamma_1(4)$ in $\tau_3$
with weight 2, with an overall S-parity equal to $(+1)$. Using the results collected in 
Appendix~\ref{secn:gamma}, matching the weights and S-parities and fixing the overall normalization in 
agreement with the perturbative contributions, we find that a solution is
\begin{equation}
\label{g023SU6}
g_{0}^{(2)}=\frac{T_6}{5} \,\frac{f_{2,+}^{(1)}\,E_4}{\big(f_{1,-}^{(3)}\big)^2}
\qquad\mbox{and}\qquad
g_{0}^{(3)}=\frac{T_6}{5} \,\frac{E_6}{f_{2,-}^{(4)}}~.
\end{equation}
By Fourier expanding the modular forms and expressing the result with bare coupling $q_0$, we do not only
recover the multi-instanton terms in (\ref{g02SU6}) and (\ref{g03SU6}) but also 
predict all other higher instanton contributions.

As before, the coefficients $g_n^{(k)}$ with $n>0$ are obtained from the coupled modular anomaly equations
(\ref{coupledeqs}), which in this case become
\begin{equation}
\begin{aligned}
\frac{\p g_1^{(k)}}{\p E_2} = \frac{5}{22}\, g_0^{(k)}\,g_0\qquad\mbox{and}\qquad
\frac{\p g_1^{(k)}}{\p \widetilde{E}_2^{(\lambda_k)}}= \frac{25}{132}\, \big(g_0^{(k)}\big)^2
\end{aligned}
\end{equation}
where the quasi-modular forms $\widetilde{E}_2^{(3)}$ for $k=2$ and $\widetilde{E}_2^{(4)}$ for $k=3$
are given in (\ref{E23tilde}) and (\ref{E24tilde}), respectively. These equations can be solved in 
a straightforward manner and the undetermined modular terms can be fixed by comparing with the 
perturbative contributions in (\ref{g12SU6}) and (\ref{g13SU6}). In this way one obtains
\begin{subequations}
\begin{align}
g_1^{(2)} &= \frac{T_6^2}{110} \left(\frac{f_{2,+}^{(1)}\,E_4^2\,E_2}{\big(f_{1,-}^{(3)}\big)^2} +
\frac{5}{6}\frac{E_4^3\,\widetilde{E}_2^{(3)}}{\big(f_{1,-}^{(3)}\big)^4}  -\frac{8}{3}\frac{f_{2,+}^{(1)}\,E_4\,E_6}{\big(f_{1,-}^{(3)}\big)^2} \right) + 
\frac{T_{12}}{22}\, \frac{f_{2,+}^{(1)}\,E_4\,E_6}{\big(f_{1,-}^{(3)}\big)^2}~,\\
g_1^{(3)} &= \frac{T_6^2}{110} \left(\frac{E_4\,E_6\,E_2}{f_{2,-}^{(4)}} +
\frac{5}{6}\frac{E_6^2\,\widetilde{E}_2^{(4)}}{\big(f_{2,-}^{(4)}\big)^2}  
-\frac{37}{12}\frac{E_6^2}{f_{2,-}^{(4)}} \right) + 
\frac{T_{12}}{22}\, \frac{E_6^2}{f_{2,-}^{(4)}}~.
\end{align}
\end{subequations}
Again, by Fourier expanding the right hand sides and expressing everything in terms of the bare 
coupling $q_0$, we retrieve the first instanton corrections in perfect agreement with the 
localization results (\ref{g3exps}) and
(\ref{g3exps1}), and predict all successive non-perturbative contributions. A similar analysis can be 
performed at the next orders; indeed we have checked that the higher coefficients $g_n^{(k)}$ are successfully
determined by these methods.

\section{Discussion and outlook}
\label{secn:concl}

In this paper we have obtained two sets of largely independent, but complementary results. 
In the first part, we calculated the period matrix for massless $\mathcal{N}=2$ SQCD theories with
gauge group SU($N$) in the massless limit in a locus of vacua possessing a $\mathbb{Z}_N$ symmetry. We 
uncovered an interesting modular structure that becomes manifest 
only when the observables are written in terms of the $\left[\frac{N}{2}\right]$ renormalized 
couplings $\tau_k$. 
In particular, we have shown that on each of these couplings, the S-duality group acts as a (generalized)
triangle group. We also proposed a non-perturbatively \emph{exact} relation between the bare coupling and the 
renormalized ones in terms of the hauptmodul of the corresponding triangle group, namely
\begin{equation}\label{UVIRgeneral2}
q_0 = \frac{ \sqrt{j_{\lambda_k}(\tau_k) -d_{\lambda_k}^{-1}} 
- \sqrt{ j_{\lambda_k}(\tau_k)\phantom{\big|}} }{ \sqrt{j_{\lambda_k}(\tau_k) -d_{\lambda_k}^{-1}} 
+ \sqrt{ j_{\lambda_k}(\tau_k)\phantom{\big|}} } ~.
\end{equation}
This relation correctly reproduces the instanton expansion and we showed that it is consistent with expectations 
from S-duality. 
While previous investigations \cite{Minahan:1997fi,Argyres:1998bn} concentrated essentially on only one of 
these effective couplings, which in our notation is $\tau_{\big[\frac{N}{2}\big]}$, our analysis shows 
that S-duality is more transparent if we consider all individual couplings $\tau_k$. Of course, we 
could select one of them and express all the others in terms of it using the exact relation (\ref{UVIRgeneral2}) via 
the bare coupling $q_0$, but then the modular structure we have described is hidden.

There are many questions that remain to be explored. For example, it would be interesting to understand from
``first principles'' the spectrum of $\lambda_k$, for which our case-by-case analysis provides
the simple answer
\begin{equation}
\lambda_k=4\sin^2\frac{k \pi}{N} \, .
\end{equation}

Using the universal formula (\ref{UVIRgeneral2}), questions about the strong coupling properties 
of the gauge theory could be addressed in an explicit way because the behaviour of the hauptmodul 
$j_{\lambda_k}$ around the strong coupling cusps in the $\tau_k$ plane is well 
understood \cite{Doran:2013npa}.

As an interesting curiosity, we observe that if we define
\begin{equation}
j^*_{\lambda_k}= -\frac{1}{4d_{\lambda_k}\,q_0}
\label{jstar}
\end{equation}
then, in the arithmetic cases (see Tab.~1) the pairs $(j_{\lambda_k}, j_{\lambda_k}^*)$ 
satisfy remarkable identities, called the Ramanujan-Sato identities, that take the form \cite{ChanCooper}:
\begin{equation}
\sum_{k=0}^\infty s_{\lambda_k}^A(k)\,\frac{1}{(j_{\lambda_k})^{k+1/2}}
= \pm \sum_{k=0}^\infty s_{\lambda_k}^B(k)\,\frac{1}{(j_{\lambda_k}^*)^{k+1/2}}~,
\end{equation}
where the $s_{\lambda_k}^{A,B}(k)$ are integers. It would be interesting to understand if these mathematical 
identities hold also in the non-arithmetic cases and if they have any interpretation within the gauge theory.

In the second part, we considered massive SQCD theories with SU($N$) gauge groups, and restricted our 
analysis to mass configurations that respect the $\mathbb{Z}_N$ symmetry of the special vacuum. We then 
showed that in this case the modular structure of the massless theory 
is deformed in an interesting manner. In particular we have proved that the period matrix 
maintains the same structure as in the massless case, while the renormalized couplings 
have a semiclassical expansion with mass dependent coefficients.
In the arithmetic theories, these coefficients are constrained by S-duality to 
satisfy coupled modular anomaly equations whose solutions are meromorphic functions of 
quasi-modular forms of the congruence subgroups of the modular group. 

A natural question to pose is whether these results can be extended to all SU$(N)$ theories. 
Since in the non-arithmetic cases the S-duality group has no subset in common with 
the modular group, we expect that the automorphic forms and Eisenstein series of the (generalized)
triangle groups should play an important role. 
This subject seems to be of recent interest in the mathematical literature \cite{Doran:2013npa} and it might be 
worthwhile to explore this possibility.

Another extension of our work would be to study the modular structure in the special vacuum with generic 
masses or in the $\Omega$-deformed theory \cite{Nekrasov:2002qd,Nekrasov:2003rj}. An incentive to study 
this problem comes from the AGT correspondence \cite{Alday:2009aq, Wyllard:2009hg}. Indeed, in the 
SU$(2)$ theory with four flavours
the non-perturbative relation between the bare and renormalized coupling plays an important role in writing the 
prepotential as quasi-modular functions. The quantum-corrected coupling constant is used to rewrite the null-
vector decoupling equation as an elliptic equation 
\cite{KashaniPoor:2012wb,Kashani-Poor:2013oza,Kashani-Poor:2014mua}. 
This in turn can be used to obtain the $\Omega$-dependent corrections to the prepotential in terms of modular 
functions
in the Nekrasov-Shatashvili limit \cite{Nekrasov:2009rc}. It would be nice to extend this approach to higher rank
gauge theories using the non-perturbative relation (\ref{UVIRgeneral2}).
In \cite{Marshakov:2009kj,Poghossian:2009mk} it was observed that for the SU(2) theory the relation between the bare and
the effective coupling is encoded in the Zamolodchikov asymptotic recursion
relation satisfied by the 4-point conformal blocks of the associated two-dimensional Liouville theory. It would
be interesting to extend this analysis to higher rank theories in order to provide further
checks on our formulas.

A more difficult but very interesting problem is to release the special vacuum constraints and analyse the theory 
at a generic point on the Coulomb moduli space, to see how the modular structures we have obtained are 
generalized. We hope to return to some of these issues in the near future.

\acknowledgments
We are indebted to Marco Bill\`o and Marialuisa Frau for collaboration and many valuable discussions, 
for sharing with us their insight and for reading the manuscript.

We would like to thank Ofer Aharony, Suresh Govindarajan, Dileep Jatkar, Renjan John, Ashoke Sen, Jan Troost, 
and especially Joseph Oesterle for useful discussions.

The results of this paper were first reported at the National Strings Meeting ($2015$), held at the Indian 
Institute for Science Education and Research, Mohali. MR would like to thank the organizers for the opportunity 
to speak at the conference and to the attendees for valuable feedback.

\appendix

\section{Quasi-modular forms}
\label{secn:gamma}
In this appendix we collect a few results on the (quasi-)modular forms of the modular group
PSL(2,$\mathbb{Z}$) and its congruence subgroups $\Gamma_1(2)$,  $\Gamma_1(3)$ and $\Gamma_1(4)$
which occur in the arithmetic theories. 
We refer to the literature
for the proofs of the various statements (see for example \cite{Koblitz,Apostol}) and only 
quote the main results that are relevant for the calculations described in the main text.

\subsection*{The modular group PSL(2,$\mathbb{Z}$) and Eisenstein series}
The Eisenstein series $E_{2n}$ are holomorphic functions of $\tau$ (with $\mathrm{Im}(\tau)\geq 0$),  
defined as
\begin{equation}
\label{defeis}
E_{2n} = \frac{1}{2\zeta(2n)}\sum_{m,n\in\mathbb{Z}^2\setminus \{0,0\}} \frac{1}{(m+n\tau)^{2n}}
\end{equation}
where $\zeta$ denotes the Riemann $\zeta$-function.
For $n>1$, the $E_{2n}$'s are modular forms of degree $2n$. In particular, under $\tau\to-1/\tau$
they transform as
\begin{equation}
\label{eissl2z}
E_{2n}\Big(\!\!-\frac{1}{\tau}\Big) = \tau^{2n} E_{2n}(\tau) = (-1)^n\big(\ii\tau\big)^{2n} E_{2n}(\tau) ~.
\end{equation}
This shows that the $S$-parity of $E_{2n}$ is $(-1)^n$.
The $E_2$ series is instead quasi-modular:
\begin{equation}
\label{eis2sl2z}
E_{2}\Big(\!\!-\frac{1}{\tau}\Big) = -\big(\ii\tau\big)^{2} \Big(E_{2}(\tau) + \frac{6}{\ii\pi\tau}\Big),
\end{equation}
and has odd $S$-parity.

All modular forms of degree $2n>6$ can be expressed in terms of $E_4$ and $E_6$; the
quasi-modular forms instead can be expressed as polynomials in $E_2$, $E_4$ and $E_6$.
The Fourier expansions of the first Eisenstein series are
\begin{equation}
\label{e246q}
\begin{aligned}
E_2 & = 1 - 24 q - 72 q^2 - 96 q^3 + \cdots~,\\
E_4 & = 1 + 240 q + 2160 q^2 + 6720 q^3 + \cdots~,\\
E_6 & = 1 - 504 q - 16632 q^2 - 122976 q^3 + \cdots
\end{aligned}
\end{equation}
where $q=\eee^{2\pi\ii\tau}$.

Let us now consider the subgroup $\Gamma'$ generated by $T$ and $S'=STS^{-1}$. As the results on the 
SU(6) theory reported in Section~\ref{secn:SU46} explicitly indicate, the following expression
\begin{equation}
f_{2,+}^{(1)}= 1 + 120 q - 6120 q^2 + 737760 q^3+\cdots 
\end{equation}
plays a crucial role in matching the modular structure of the period matrix with the multi-instanton calculations.
We notice that this expansion is accounted for if we write
\begin{equation}
f_{2,+}^{(1)}= \big(E_4\big)^{\frac{1}{2}}~.
\end{equation}
The presence of the square root seems to suggest that the modular group can be viewed as a two-sheeted
cover of $\Gamma'$. 
Moreover we observe that everything is consistent by requiring that $f_{2,+}^{(1)}$ be a 
modular form of weight 2 under $\Gamma'$ and with positive $S$-parity. This also explains the notation 
we have used.

\subsection*{The congruence subgroup $\Gamma_1 (2)$}
To construct the modular forms of $\Gamma_1 (2)$ we first define the following functions
\begin{equation}
f_{4,\pm}^{(2)}(\tau) = \left(\frac{\eta^2(\tau)}{\eta(2\tau)} \right)^8 
\pm 64 \left(\frac{\eta^2(2\tau)}{\eta(\tau)} \right)^8
\end{equation}
where $\eta(\tau)$ is the Dedekind $\eta$-function. Their Fourier expansions are
\begin{equation}
\begin{aligned}
f_{4,+}^{(2)}&= 1+48 q+624 q^2+1344 q^3+\cdots ~,\\
f_{4,-}^{(2)}&=1-80 q-400 q^2-2240 q^3+\cdots~,
\end{aligned}
\end{equation}
where as usual $q=\eee^{2\pi\ii\tau}$.
These functions are modular forms $\Gamma_1(2)$ of weight 4 \cite{Koblitz,Apostol}, as evinced by 
their behavior under the $S'$-transformation:
\begin{equation}
f_{4,\pm}^{(2)}\left(\frac{\tau}{1-2\tau}\right) = (1-2\tau)^4 f_{4,\pm}^{(2)}(\tau)~.
\end{equation}
In addition, using the modular transformation properties of the Dedekind $\eta$-function and in particular
\begin{equation}
\eta\Big(\!\!-\frac{1}{\tau}\Big) = \sqrt{-\ii\tau}\,\eta(\tau)~,
\end{equation}
one can easily check that 
\begin{equation}
f_{4,\pm}^{(2)}\Big(\!\!-\frac{1}{2\tau}\Big) = \pm \big(\ii \sqrt{2}\tau\big)^4 \,f_{4,\pm}^{(2)}(\tau)~.
\end{equation}
Thus $f_{4,\pm}^{(2)}$ have weight 4 and $S$-parity $(+1)$ and $(-1)$ respectively, as the notation itself 
suggests.

Now consider the square-root of $f_{4,+}^{(2)}$, namely
\begin{equation}
f_{2,+}^{(2)} := \big(f_{4,+}^{(2)}\big)^{\frac{1}{2}} = 1 + 24q+24q^2+96q^3 + \cdots~.
\end{equation}
This is a modular form of $\Gamma_1(2)$ with weight 2 and positive $S$-parity. Indeed,
\begin{equation}\label{Sonf2}
f_{2,+}^{(2)}\Big(\!\!-\frac{1}{2\tau}\Big) = \big(\ii\sqrt{2}\tau\big)^2 \,f_{2,+}^{(2)}(\tau) ~.
\end{equation}
The modular forms of $\Gamma_1(2)$ form a ring generated by $f_{2,+}^{(2)}$ 
and $f_{4,-}^{(2)}$.

In order to study quasi-modular forms of $\Gamma_1(2)$ let us consider the second Eisenstein series $E_2$
which satisfies
\begin{equation}
\begin{aligned}
E_2 \Big(\!\!-\frac{1}{2\tau}\Big) &= \big(4\tau^2\big)\,E_2(2\tau) + \frac{12\tau}{\ii\pi}~,\\
E_2(2\tau) &= \frac{1}{2}\,E_2(\tau)+\frac{1}{2}\,f_{2,+}^{(2)}(\tau)~.
\end{aligned}
\end{equation}
These equations naturally lead us to introduce the following combination
\begin{equation}
\widetilde{E}_2^{(2)}= E_2+\frac{1}{2}f_{2,+}^{(2)} ~=~ \frac{3}{2}-12 q-60 q^2-48 q^3+\cdots~.
\label{E22tilde}
\end{equation}
Using \eqref{Sonf2}, it is easy to check that 
\begin{equation}
\widetilde{E}_2^{(2)} \Big(\!\!-\frac{1}{2\tau}\Big) = -\big(\sqrt{2}\ii\tau\big)^2\,
\Big(\widetilde{E}_2^{(2)}(\tau) + \frac{6}{\ii\pi\tau}\Big)~,
\end{equation}
which shows that $\widetilde{E}_2^{(2)}$ transforms under S-duality similarly to $E_2$ and has negative
$S$-parity.

\subsection*{The congruence subgroup $\Gamma_1 (3)$}
To construct the modular forms of $\Gamma_1 (3)$ we first define the following functions
\begin{equation}
f_{3,\pm}^{(3)}(\tau) = \left(\frac{\eta^3(\tau)}{\eta(3\tau)} \right)^3 
\mp 27 \left(\frac{\eta^3(3\tau)}{\eta(\tau)} \right)^3
\end{equation}
whose Fourier expansions are
\begin{equation}
\begin{aligned}
f_{3,+}^{(3)}&= 1 - 36 q - 54 q^2 - 252 q^3+\cdots ~,\\
f_{3,-}^{(3)}&= 1+18 q+108 q^2+234 q^3+\cdots~,
\end{aligned}
\end{equation}
where as usual $q=\eee^{2\pi\ii\tau}$.
These functions are modular forms $\Gamma_1(3)$ of weight 3 \cite{Koblitz,Apostol}. Under the $S'$-
transformation, they behave as 
\begin{equation}
f_{3,\pm}^{(3)}\left(\frac{\tau}{1-3\tau}\right) = (1-3\tau)^3 f_{3,\pm}^{(3)}(\tau)~,
\end{equation}
while under the $S$-transformation they change as 
\begin{equation}
f_{3,\pm}^{(3)}\Big(\!\!-\frac{1}{3\tau}\Big) = \pm \big(\ii \sqrt{3}\tau\big)^3 \,f_{3,\pm}^{(3)}(\tau)~,
\end{equation}
as one can easily check using the modular properties of the Dedekind function. The last equation shows that
$f_{3,\pm}^{(3)}$ have $S$-parity $(+1)$ and $(-1)$, respectively, as also the notation suggests.

Now consider the cube-root of $f_{3,-}^{(3)}$, namely
\begin{equation}
f_{1,-}^{(3)} := \big(f_{3,-}^{(3)}\big)^{\frac{1}{3}} = 1+6q+6q^3 + \cdots~.
\end{equation}
This is a modular form of $\Gamma_1(3)$ with weight 1 and negative $S$-parity. Indeed,
\begin{equation}\label{Sonf3}
f_{1,-}^{(3)}\Big(\!\!-\frac{1}{3\tau}\Big) = -\big(\ii\sqrt{3}\tau\big) \,f_{1,-}^{(3)}(\tau) ~.
\end{equation}
The modular forms of $\Gamma_1(3)$ form a ring generated by $f_{1,-}^{(3)}$ 
and $f_{3,+}^{(3)}$.

In order to study quasi-modular forms of $\Gamma_1(3)$ we have to consider the second Eisenstein series 
$E_2$ which satisfies
\begin{equation}
\begin{aligned}
E_2 \Big(\!\!-\frac{1}{3\tau}\Big) &= \big(9\tau^2\big)\,E_2(3\tau) + \frac{18\tau}{\ii\pi}~,\\
E_2(3\tau) &= \frac{1}{3}\,E_2(\tau)+\frac{2}{3}\,\big(f_{1,-}^{(3)}(\tau)\big)^2~.
\end{aligned}
\end{equation}
These equations naturally lead us to introduce the following combination
\begin{equation}
\widetilde{E}_2^{(3)}= E_2+\big(f_{1,-}^{(3)}\big)^2 ~=~ 2 - 12 q - 36 q^2 - 84 q^3+\cdots~.
\label{E23tilde}
\end{equation}
Using \eqref{Sonf3}, it is easy to check that 
\begin{equation}
\widetilde{E}_2^{(3)} \Big(\!\!-\frac{1}{3\tau}\Big) = -\big(\sqrt{3}\ii\tau\big)^2\,
\Big(\widetilde{E}_2^{(3)}(\tau) + \frac{6}{\ii\pi\tau}\Big)~,
\end{equation}
which shows that $\widetilde{E}_2^{(3)}$ transforms under S-duality similarly to $E_2$ and has negative
$S$-parity.

\subsection*{The congruence subgroup $\Gamma_1 (4)$}

The ring of modular forms of $\Gamma_1(4)$ is generated by the weight-2 modular forms which we denote 
$f_{2,\pm}^{(4)}$. They are defined as 
\begin{equation}
\begin{aligned}
f_{2,+}^{(4)}(\tau)&:= \theta_3^4(2\tau)~ =~ 1+8 q+24 q^2+32 q^3 + \cdots~, \\
f_{2,-}^{(4)}(\tau)&:= \theta_4^4(2\tau)-\theta_2^4(2\tau) ~=~
1-24q+24q^2-96q^3+\cdots~,
\end{aligned}
\end{equation}
where the $\theta_a$'s are the standard Jacobi $\theta$-functions and as usual $q=\eee^{2\pi\ii\tau}$.
Using the modular properties of the $\theta$-functions and in particular
\begin{equation}
\theta_2\Big(\!\!-\frac{1}{\tau}\Big)=\sqrt{-\ii\tau}\,\theta_4(\tau)~,
\quad
\theta_3\Big(\!\!-\frac{1}{\tau}\Big)=\sqrt{-\ii\tau}\,\theta_3(\tau)~,
\quad
\theta_4\Big(\!\!-\frac{1}{\tau}\Big)=\sqrt{-\ii\tau}\,\theta_2(\tau)~,
\end{equation}
it is easy to show that
\begin{equation}
f^{(4)}_{2, \pm}\Big(\!\!-\frac{1}{4\tau}\Big) = \pm (2\ii \tau)^2\,f^{(4)}_{2, \pm}(\tau)~.
\end{equation}
Thus $f^{(4)}_{2, \pm}$ have $S$-parity $(+1)$ and $(-1)$ respectively.

In order to study quasi-modular forms of $\Gamma_1(4)$ we have to consider the second Eisenstein series 
$E_2$ which satisfies
\begin{equation}
\begin{aligned}
E_2 \Big(\!\!-\frac{1}{4\tau}\Big) &= \big(4\tau\big)^2\,E_2( 4\tau) 
+ \frac{24 \tau}{\ii \pi} ~,\\
E_2( 4\tau) &= \frac{1}{4}\, E_2(\tau) + \frac{3}{4}\, f_{2,+}^{(4)}(\tau)~.
\end{aligned}
\end{equation}
These equations suggest to introduce the following combination
\begin{equation}
\widetilde{E}_2^{(4)} = E_2 + \frac{3}{2}\, f_{2,+}^{(4)}~=~\frac{5}{2} - 12 q - 36 q^2 - 48 q^3+\cdots~,
\label{E24tilde} 
\end{equation}
which under S-duality transforms in a way similar to $E_2$, namely
\begin{equation}
\widetilde{E}_2^{(4)} \Big(\!\!-\frac{1}{4\tau} \Big) = -\big( 2\ii\tau \big)^2 \,\Big( 
\widetilde{E}_2^{(4)} + \frac{6}{\ii \pi \tau} \Big) ~.
\end{equation}
This equation shows that $\widetilde{E}_2^{(4)}$ is a quasi-modular form with weight 2 and negative 
$S$-parity.

\section{\boldmath $N=7$}
\label{secn:N7}
In this appendix we briefly report the results for the massless SU$(7)$ SQCD theory in the special vacuum.

The quantum corrected period matrix takes the form
\begin{equation}
\Omega=\tau_1\,\mathcal{M}_1+\tau_2\,\mathcal{M}_2+\tau_3\,\mathcal{M}_4
\label{OmegaSU7}
\end{equation}
where
\begin{equation}
\mathcal{M}_k = \sum_{\ell=1}^3\lambda_{k\ell}\,\mathcal{G}_\ell
\end{equation}
with
\begin{equation}
\lambda_k= 4\sin^2\frac{k\pi}{7}\,=\,4\cos^2\frac{(7-2k)\pi}{14}
\end{equation}
and
\begin{equation}
\mathcal{G}_1=\begin{pmatrix}
2& 0& 1& 1& 1& 2\\ 
0& 0& -1& 0& 0& 1\\ 
1& -1& 0& -1& 0& 1\\ 
1& 0& -1& 0& -1& 1\\ 
1& 0& 0& -1& 0& 0\\ 
2& 1& 1& 1& 0& 2
\end{pmatrix}\,~
\mathcal{G}_2=\begin{pmatrix}
0& 1& -1& 0& 1& -1\\ 
1& 2& 1& 0& 2& 1\\ 
-1& 1& 0& 0& 0& 0\\ 
0& 0& 0& 0& 1& -1\\ 
1& 2& 0& 1& 2& 1\\ 
-1& 1& 0& -1& 1& 0
\end{pmatrix}\,~
\mathcal{G}_3=
\begin{pmatrix}
0& 0& 1& 0& -1& 0\\ 
0& 0& 1& 1& -1& -1\\
1& 1& 2& 2& 1& 0\\
0& 1& 2& 2& 1& 1\\
-1& -1& 1& 1& 0& 0\\
0& -1& 0& 1& 0& 0
\end{pmatrix}~.
\end{equation}
These are specific examples of the matrices defined through Eq. (\ref{Gk}) of the main text. Finally, up to two 
instantons we find that the three renormalized couplings are given by
\begin{equation}
\label{su7nekrasov}
\begin{aligned}
2\pi\ii\,\tau_1 &= \log q_0 + \ii \pi + \log\big(4d_{\lambda_1}\big)  + \frac{12}{49} q_0 
+ \frac{192}{2401}q_0^2 +\cdots~,\\ 
2\pi\ii\,\tau_2 &= \log q_0 + \ii \pi + \log\big(4d_{\lambda_2}\big) + \frac{20}{49} q_0 
+ \frac{370}{2401}q_0^2 +\cdots ~,\\
2\pi\ii\,\tau_3 &= \log q_0 + \ii \pi + \log\big(4d_{\lambda_3}\big)   + \frac{24}{49} q_0 
+ \frac{474}{2401}q_0^2 +\cdots ~,
\end{aligned}
\end{equation}
with
\begin{equation}
d_{\lambda_1}=4611.1803\cdots~,\quad d_{\lambda_2}=
163.6225\cdots~,\quad d_{\lambda_3}=69.8572\cdots~.
\label{dlambdas}
\end{equation}
It is worth noticing that all coefficients in the instanton expansion of the renormalized couplings are rational.
According to the general discussion of Section~\ref{secn_triangle}, these formulas should follow upon using the
hauptmoduln of certain (generalized) triangle groups in the universal formula (\ref{UVIRgeneral}). We now show 
that this is indeed the case for the SU(7) theory.

Let us start from $k=3$. Here we have $\lambda_3 = 4\cos^2\frac{\pi}{14}$ and thus the S-duality group is 
simply the Hecke group H(14) whose type is $\mathbf{t}=(2,14,\infty)$. Applying the formulas
of Section~\ref{subsecn:triangle}, it is not difficult to find that the corresponding hauptmodul is
\begin{equation}
j_{\lambda_3} = \frac{1}{q_3} + \frac{37}{98}\frac{1}{d_{\lambda_3}} 
+ \frac{2587}{38416} \frac{q_3}{d_{\lambda_3}^2} 
+ \frac{899}{117649}\frac{q_3^2}{ d_{\lambda_3}^3} +\cdots 
\end{equation}
where $d_{\lambda_3}$ is precisely the same number given in (\ref{dlambdas}).

Now let us put $k=2$. In this case we have $\lambda_2 = 4\cos^2\frac{3\pi}{14}$ which implies that the
S-duality group is a generalized triangle group with type $t=\big(2,\frac{14}{3},\infty\big)$.
As we observed in the main text, the formulas for the hauptmoduln can be formally extended also when the 
type has a rational entry. In this case we find
\begin{equation}
j_{\lambda_2} =\frac{1}{q_2} +\frac{ 39}{98}\frac{1}{d_{\lambda_2}}  
+ \frac{2571}{38416} \frac{q_2}{d_{\lambda_2}^2}  
+ \frac{4435}{705894} \frac{q_2^2}{d_{\lambda_2}^3}  +\cdots
\end{equation}
where $d_{\lambda_2}$ is exactly as in (\ref{dlambdas}).

Finally for $k=1$, we have $\lambda_1 = 4\cos^2\frac{5\pi}{14}$ leading to a generalized triangle group
with type $t=\big(2,\frac{14}{5},\infty\big)$. In this case the corresponding hauptmodul is
\begin{equation}
j_{\lambda_1} =\frac{1}{q_1} + \frac{43}{98} \frac{1}{d_{\lambda_1}}
+ \frac{2521}{38416 }\frac{q_1}{d_{\lambda_1}^2} 
+ \frac{2573}{705894} \frac{q_1^2}{d_{\lambda_1}^3}  +\cdots
\end{equation}
with $d_{\lambda_1}$ given in (\ref{dlambdas}).

If we now plug these expansions in the universal formula (\ref{UVIRgeneral}) and invert the resulting series, we 
perfectly match the instanton results (\ref{su7nekrasov}) obtained from localization, thus confirming also in this 
case the consistency of our proposal.

\providecommand{\href}[2]{#2}\begingroup\raggedright\endgroup
\end{document}